\documentclass{JHEP3} 
\usepackage{epsfig}
\usepackage{amsbsy}
\usepackage{amsfonts}
\usepackage{amssymb}

\title{Bayesian View Of Solar Neutrino Oscillations}

\author{M.V. Garzelli and C. Giunti
\\
INFN, Sezione di Torino,
\\
and
\\
Dipartimento di Fisica Teorica,
Universit\`a di Torino,
\\
Via P. Giuria 1, I--10125 Torino, Italy
\\
E-mail: \email{garzelli@to.infn.it}, \email{giunti@to.infn.it}}

\received{\today}
\accepted{\today}
\preprint{\hepph{0108191}}

\abstract{
We present the results of a
Bayesian analysis
of solar neutrino data
in terms of $\nu_e\to\nu_{\mu,\tau}$
and
$\nu_e\to\nu_s$
oscillations,
where $\nu_s$ is a sterile neutrino.
We perform a Rates Analysis of the rates of solar neutrino experiments,
including the first SNO CC result,
and spectral data of the CHOOZ experiment,
and a Global Analysis that takes into account also
the Super-Ka\-mio\-kan\-de day and night electron energy spectra.
We show that
the Bayesian analysis of solar neutrino data
does not suffer any problem from the inclusion
of the numerous bins of the CHOOZ and Super-Ka\-mio\-kan\-de
electron energy spectra
and
allows to
reach the same conclusions
on the favored type of neutrino transitions
and on the determination of the most favored
values of the oscillation parameters
in both the Rates and Global Analysis.
Our Bayesian analysis shows that
$\nu_e\to\nu_s$ transitions
are strongly disfavored with respect to
$\nu_e\to\nu_{\mu,\tau}$ transitions.
In the case of $\nu_e\to\nu_{\mu,\tau}$ oscillations,
the Large Mixing Angle region
is favored by the data
(86\% probability),
the LOW region has some small chance
(13\% probability),
the Vacuum Oscillation region
is almost excluded
(1\% probability)
and
the Small Mixing Angle region is practically excluded
(0.01\% probability).
We calculate also
the marginal
posterior probability distributions
for
$\tan^2\!\vartheta$
and
$\Delta{m}^2$
in the case of $\nu_e\to\nu_{\mu,\tau}$ oscillations
and we show that
the data imply large mixing almost with certainty
and large values of
$\Delta{m}^2$
are favored
($
2 \times 10^{-6} \, \mathrm{eV}^2
<
\Delta{m}^2
<
10^{-3} \, \mathrm{eV}^2
$
with 86\% probability).
We present also
the results of a standard least-squares
analysis of solar neutrino data
and we show that the standard goodness of fit test
is not able to reject pure
$\nu_e\to\nu_s$ transitions.
The likelihood ratio test,
which is insensitive to the number of bins of the CHOOZ and Super-Ka\-mio\-kan\-de
energy spectra,
allows to reject pure
$\nu_e\to\nu_s$ transitions
in favor of
$\nu_e\to\nu_{\mu,\tau}$ transitions
only in the Global Analysis.
}

\keywords{Solar Neutrinos, Neutrino Physics, Statistical Methods}

\begin{document}

\section{Introduction}
\label{Introduction}

The experimental investigation of the solar neutrino problem
(see \cite{Bahcall:2000ue})
has received recently a fundamental contribution
from the first data of the
SNO solar neutrino experiment
\cite{SNO-01},
in which charged-current interactions
of solar $^8\mathrm{B}$ electron neutrinos
with deuterium
have been observed with a rate about 0.35 times of
that predicted by the BP2000 Standard Solar Model (SSM) \cite{BP2000}.
A model independent comparison of SNO and Super-Ka\-mio\-kan\-de
\cite{SK-sun-01} data
show that $\nu_\mu$ or $\nu_\tau$
are present in the flux of solar neutrino on Earth
at $3.06\sigma$
\cite{SNO-01,Fogli:2001vr}
(see Ref.~\cite{Giunti-aoe-01} for a discussion on the
probability content of this statement).
Since the simplest explanation
of the presence of $\nu_\mu$ or $\nu_\tau$
in the flux of solar neutrino on Earth
is neutrino oscillations due to neutrino masses and mixing
(see \cite{BGG-review-98}),
this is the second model-independent
evidence in favor of neutrino mixing,
after the up-down asymmetry of
multi-GeV atmospheric $\nu_\mu$-induced events
discovered in the Super-Ka\-mio\-kan\-de experiment
\cite{SK-evidence-98}.

In the simplest case of mixing of two neutrinos
$\nu_e$ and $\nu_x$
that we consider here,
we have
\begin{eqnarray}
\null & \null & \null
\nu_e = \cos\vartheta \, \nu_1 + \sin\vartheta \, \nu_2
\,,
\nonumber
\\
\null & \null & \null
\nu_x = -\sin\vartheta \, \nu_1 + \cos\vartheta \, \nu_2
\,,
\label{mixing}
\end{eqnarray}
where $\vartheta$ is the mixing angle
and
$\nu_1$, $\nu_2$
are massive neutrinos
with masses
$m_1$, $m_2$,
respectively.
The flavor neutrino $\nu_x$
could be either active, $x=a$ with $a=\mu,\tau$,
or sterile, $x=s$
(see \cite{BGG-review-98}).

Following the release of the SNO first results,
the data of solar neutrino experiments
(Homestake \cite{Homestake-98},
GALLEX \cite{GALLEX-99},
SAGE \cite{SAGE-nu00},
GNO \cite{GNO-00},
Super-Ka\-mio\-kan\-de \cite{SK-sun-01},
SNO \cite{SNO-01})
have been analyzed in terms of neutrino oscillations
in Refs.~\cite{Barger:2001zs,Fogli:2001vr,Bahcall:2001zu,%
Bandyopadhyay:2001aa,Creminelli:2001ij,%
Berezinsky-Lissia-01,Berezinsky:2001se,%
Krastev:2001tv}.
The above-mentioned $3.06\sigma$
model-independent evidence in favor of the presence of
$\nu_\mu$ or $\nu_\tau$
in the flux of solar neutrino on Earth
clearly implies that the solution of the solar neutrino problem
in terms of oscillations of solar $\nu_e$'s
into active $\nu_\mu$ or $\nu_\tau$
is strongly
favored with respect to transitions into sterile neutrinos $\nu_s$.
Therefore,
assuming the BP2000 SSM prediction
for the neutrino fluxes produced in the core of the Sun \cite{BP2000}
and using a standard least-squares analysis,
the authors
of Refs.~\cite{Fogli:2001vr,Bahcall:2001zu,%
Bandyopadhyay:2001aa,Krastev:2001tv}
calculated the allowed regions in the plane of the oscillation parameters
$\tan^2\!\vartheta$ and $\Delta{m}^2$
in the case of $\nu_e \to \nu_a$ transitions
($\Delta{m}^2 \equiv m_2^2 - m_1^2$)\footnote{
The authors of Ref.~\cite{Bahcall:2001zu}
took also into account the very interesting possibility of simultaneous
$\nu_e \to \nu_a$
and
$\nu_e \to \nu_s$
transitions,
that is allowed in four-neutrino mixing schemes
(see
\cite{BGG-review-98,DGKK-99,Concha-foursolar-00,%
Giunti-Laveder-3+1-00,Gonzalez-Garcia:2001uy}
and references therein).
In this case,
besides $\tan^2\!\vartheta$ and $\Delta{m}^2$,
there is an additional parameter $\cos^2\eta$ that regulates the
relative amount of
$\nu_e \to \nu_a$
and
$\nu_e \to \nu_s$
transitions
($0\leq\eta\leq\pi/2$).
We do not consider here this possibility
because of its computational difficulties
(Bayesian credible regions in the
$\tan^2\!\vartheta$--$\Delta{m}^2$ plane
must be calculated integrating the posterior probability distribution function
for the oscillation parameters over $\cos^2\eta$).
}.

The standard least-squares method for the analysis of
solar neutrino data is an approximation of
the rigorous frequentist method to calculate
allowed regions with exact coverage
\cite{Garzelli-Giunti-sf-00,%
Creminelli:2001ij,%
Garzelli-Giunti-ugs-01}.
It has been shown in
Refs.~\cite{Garzelli-Giunti-sf-00,%
Garzelli-Giunti-ugs-01}
that an analysis of
solar neutrino data
with a rigorous frequentist method
may lead to results that differ significantly from those
obtained with the standard least-squares method.
Unfortunately, the implementation of a
rigorous frequentist method in the analysis of solar neutrino data
is complicated and computer-time consuming
because it is necessary to calculate the
distribution probability of the estimator of the parameters
for all possible values of the parameters.
Moreover,
several rigorous frequentist methods
are available and it is not clear which is the
most appropriate (if there is one)
for the analysis of solar neutrino data
(see Refs.~\cite{Garzelli-Giunti-sf-00,%
Creminelli:2001ij,%
Garzelli-Giunti-ugs-01} and references therein).

In this paper we present the results of an
analysis of solar neutrino data
in terms of two-neutrino oscillations
in the framework of
Bayesian Probability Theory
(see also \cite{Creminelli:2001ij}).
This analysis has practically the same difficulties
of the standard least-squares analysis
and can therefore be easily performed by the groups
specialized in the analysis of solar neutrino data.

The Bayesian analysis of solar neutrino data is not
approximate as the least-squares analysis in the framework
of Frequentist Statistics.
Furthermore,
Bayesian Probability Theory has several advantages over
Frequentist Statistics,
that have been discussed in several books and papers
(see \cite{Jeffreys-book-39,Loredo-90,Loredo-92,Jaynes-book-95,D'Agostini-99}).
Here we notice only a few facts:
\begin{enumerate}
\item \label{1}
All human statements,
including scientific ones,
represent knowledge (belief) with some degree of uncertainty.
Bayesian Probability Theory
allows to quantify this uncertainty
through the natural definition of probability
as degree of belief.
\item \label{2}
Bayesian Probability Theory
allows to calculate
through Bayes Theorem
the improvement of knowledge
as a consequence of experimental measurements.
This is how our mind works and how science improves.
Therefore,
Bayesian Probability Theory is the natural statistical tool for scientists.
\item \label{3}
Probability of an event in Frequentist Statistics
is defined as the asymptotic relative frequency of occurrence of the event
in a large set of experiments.
Obviously such set is never available in practice.
Therefore,
Frequentist Statistics
is based on imaginary data.
On the other hand,
all inferences in Bayesian Probability Theory
are based only on the data that actually occurred.
\item \label{4}
Because of the definition of probability
in Frequentist Statistics,
the results obtained have usually good properties
(\textit{i.e.} they are reliable)
only in the case of large data sets.
In frontier research
often a small number of experimental data are available
(as in the case of solar neutrinos)
and the scientist is not interested in long-term behavior of inferences,
but in getting the best possible inference from the available data.
Bayesian Probability Theory
satisfies this wish.
\item \label{5}
Results in
Frequentist Statistics
are based on hypothetical data,
\textit{i.e.} data that could have been observed but did not occur.
Often different scientists may have different
ideas on which are the relevant hypothetical data,
leading to different conclusions
(for example, the so-called ``optional stopping problem''
is well known and discussed in the literature).
\item \label{6}
In Frequentist Statistics
there is no way to take into account theoretical and systematic errors,
that are not random variables.
Nevertheless,
since the great majority of scientific measurements
suffer of systematic errors
and scientific inferences
depend on theoretical errors
(they are both crucial in the analysis of solar neutrino data),
frequentists treat these errors
as if they were random variables.
Nobody knows the meaning of results obtained in this way,
with an inconsistent method.
Bayesian Probability Theory
obviously can treat theoretical and systematic errors
on the same footing as statistical ones,
leading to consistent results.
\item \label{7}
Once a problem is well posed the application of
Bayesian Probability Theory is clear and straightforward
and leads to unique results.
On the other hand,
in the framework of Frequentist Statistics
several arbitrary and difficult choices
that can lead to different results
are necessary.
That is why popular methods (as least-squares)
that sometimes have poor performances
are widely used without a real understanding of the motivations.
\item \label{8}
Since
Bayesian Probability Theory
allows to calculate the improvement of knowledge
as a consequence of physical observations,
a prior knowledge is necessary.
It has been argued by
advocates of Frequentist Statistics
that prior knowledge
is subjective,
leading to undesirable subjectivity
in the derivation of scientific results
(forgetting the often less clear subjective choices
of method, estimator, etc.
necessary in Frequentist Statistics).
On this problem widely discussed in the literature
we want only to remark that all human activities,
including scientific research,
have some degree of subjectivity,
but communication and collaboration among
people working in a field allow to
reach an agreement on the most reasonable prior
knowledge in the field and a way to quantify it.
Once the prior
knowledge is fixed,
Bayesian Probability Theory
leads to unique conclusions.
\end{enumerate}

The plan of the paper is as follows.
In Section~\ref{chi2}
we perform a standard least-squares
analysis of solar neutrino data.
From the calculational point of view
this is not an additional task,
because
the calculation of the least-squares function is a necessary step
in order to derive the Bayesian probability distribution
for the parameters.
We consider the two cases of $\nu_e\to\nu_a$ transitions (Active)\footnote{
The treatment of solar $\nu_e\to\nu_a$ transitions
in the framework of the simple two-neutrino mixing scheme in Eq.~(\ref{mixing})
with $\nu_x=\nu_a$
is valid with good approximation in the case of mixing
of three neutrinos
($\nu_e$, $\nu_\mu$, $\nu_\tau$)
with small $U_{e3}$
\cite{Bilenky-Giunti-CHOOZ-98,Fogli-Lisi-Montanino-Palazzo-sun-3nu-00},
as indicated by the results of the CHOOZ
long-baseline $\bar\nu_e$ disappearance experiment
\cite{Apollonio:1999ae}.
In this case $\nu_a$ is a linear combination
of $\nu_\mu$ and $\nu_\tau$.
Furthermore,
the treatment is also valid in 3+1 four-neutrino mixing schemes
in which
the sterile neutrino
is practically decoupled from the active ones
($1-|U_{s4}|^2 \ll 1$)
\cite{Barger-Fate-2000}
and
$U_{e3}$
is small, as indicated by CHOOZ data
\cite{Giunti-Laveder-3+1-00}.
},
and
$\nu_e\to\nu_s$ transitions (Sterile).
In Section~\ref{Bayesian}
we present our
Bayesian analysis of solar neutrino data.
In Section~\ref{Active_Or_Sterile}
we compare the probabilities of
the models with $\nu_e\to\nu_a$
and $\nu_e\to\nu_s$
transitions.
In Section~\ref{Credible}
we present the results of our calculation of
Bayesian allowed regions (credible regions)
in the $\tan^2\!\vartheta$--$\Delta{m}^2$
plane for Active transitions.
In Section~\ref{Marginal}
we marginalize the posterior Bayesian probability distribution
for Active transitions
in order to calculate the posterior Bayesian probability distributions
for $\tan^2\!\vartheta$ and $\Delta{m}^2$.
Finally, in Section~\ref{Conclusions}
we draw our conclusions.

In both the standard least-squares and Bayesian analyses
we consider first only the total rates
measured by solar neutrino experiments
(Rates Analysis)
and then the total rates of the
Homestake,
GALLEX+GNO+SAGE,
SNO experiments
and
the Super-Ka\-mio\-kan\-de day and night electron energy spectra
(Global Analysis).
In both the Rates Analysis and the Global Analysis
we consider also the data of the CHOOZ experiment,
that are important
because they exclude large mixing for
$\Delta{m}^2 \gtrsim 10^{-3} \, \mathrm{eV}^2$
\cite{Apollonio:1999ae}.
Therefore,
we consider only values of
$\Delta{m}^2$
smaller than $10^{-3} \, \mathrm{eV}^2$
and we will see that, consistently,
the allowed regions for the oscillation parameters
are limited below this value
of $\Delta{m}^2$.

In this paper we use the standard names
for the regions in
$\tan^2\!\vartheta$--$\Delta{m}^2$ plane
(see \cite{BGG-review-98}):
Small Mixing Angle (SMA) for
\begin{equation}
10^{-4}
<
\tan^2\!\vartheta
<
10^{-1}
\,,
\quad
10^{-8} \, \mathrm{eV}^2
<
\Delta{m}^2
<
10^{-3} \, \mathrm{eV}^2
\,,
\label{SMA}
\end{equation}
Large Mixing Angle (LMA) for
\begin{equation}
10^{-1}
<
\tan^2\!\vartheta
<
10
\,,
\quad
2 \times 10^{-6} \, \mathrm{eV}^2
<
\Delta{m}^2
<
10^{-3} \, \mathrm{eV}^2
\,,
\label{LMA}
\end{equation}
LOW for
\begin{equation}
10^{-1}
<
\tan^2\!\vartheta
<
10
\,,
\quad
10^{-8} \, \mathrm{eV}^2
<
\Delta{m}^2
<
2 \times 10^{-6} \, \mathrm{eV}^2
\,,
\label{LOW}
\end{equation}
Vacuum Oscillation (VO) for
\begin{equation}
10^{-1}
<
\tan^2\!\vartheta
<
10
\,,
\quad
10^{-11} \, \mathrm{eV}^2
<
\Delta{m}^2
<
10^{-8} \, \mathrm{eV}^2
\,.
\label{VO}
\end{equation}

\section{Standard Least-Squares Analysis}
\label{chi2}

In our
standard least-squares analysis (often called ``$\chi^2$ analysis'')
we consider first in the Rates Analysis of
Section~\ref{chi2-rates}
only the total rates
measured by solar neutrino experiments (see Table~\ref{rates}),
and the CHOOZ positron spectra given in Ref.~\cite{Apollonio:1999ae}.
Then
in the Global Analysis of Section~\ref{chi2-global}
we consider also
the Super-Ka\-mio\-kan\-de data on the day and night electron energy spectra.

\TABULAR[t]{|c|c|}{
\hline
\vphantom{$\Big|$}
Detection Material and Process
&
Data
\\
\hline
\vphantom{$\Big|$}
\begin{tabular}{c}
${}^{37}\mathrm{Cl}$:
\quad
$\nu_e + {}^{37}\mathrm{Cl} \to {}^{37}\mathrm{Ar} + e^-$
\\
(Homestake \protect\cite{Homestake-98})
\end{tabular}
&
$ 2.56 \pm 0.23 \, \mathrm{SNU} $
\\
\hline
\vphantom{$\Big|$}
\begin{tabular}{c}
${}^{71}\mathrm{Ga}$:
\quad
$\nu_e + {}^{71}\mathrm{Ga} \to {}^{71}\mathrm{Ge} + e^-$
\\
(GALLEX \protect\cite{GALLEX-99}
+
GNO \protect\cite{GNO-00}
+
SAGE \protect\cite{SAGE-nu00})
\end{tabular}
&
$ 74.7 \pm 5.1 \, \mathrm{SNU} $
\\
\hline
\vphantom{$\Big|$}
\begin{tabular}{c}
$\mathrm{D}_2\mathrm{O}$:
\quad
$\nu_e + d \to p + p + e^-$
\\
(SNO \protect\cite{SNO-01})
\end{tabular}
&
$ 0.347 \pm 0.028 $
\\
\hline
\vphantom{$\Big|$}
\begin{tabular}{c}
$\mathrm{H}_2\mathrm{O}$:
\quad
$\nu + e^- \to \nu + e^-$
\\
(Super-Ka\-mio\-kan\-de \protect\cite{SK-sun-01})
\end{tabular}
&
$ 0.459 \pm 0.017 $
\\
\hline
}
{
\label{rates}
The rates measured in solar neutrino experiments.
The rates of the Homestake and GALLEX+SAGE+GNO
experiments are expressed in SNU units
($ 1 \, \mathrm{SNU} \equiv 10^{-36} \,
\mathrm{events} \, \mathrm{atom}^{-1} \, \mathrm{s}^{-1} $),
whereas the results of the Ka\-mio\-kan\-de and SNO experiments
are expressed in terms of the ratio of the
experimental rate and the BP2000 Standard Solar Model
prediction \protect\cite{BP2000}.
The statistical and systematic uncertainties have been added in quadrature.
The GALLEX+SAGE+GNO rate is a weighted average of the
GALLEX+GNO rate reported in Ref.~\protect\cite{GNO-00}
and the SAGE rate reported in Ref.~\protect\cite{SAGE-nu00}.
The rate of the SNO experiment is that measured through CC weak interactions.
}

\subsection{Rates Analysis}
\label{chi2-rates}

Our least-squares function for the Rates Analysis is written as
\begin{equation}
X^2
=
X^2_{\mathrm{S}}
+
X^2_{\mathrm{C}}
\,,
\label{X2}
\end{equation}
where
$X^2_{\mathrm{S}}$
takes into account the rates measured in solar neutrino experiments
and
$X^2_{\mathrm{C}}$
takes into account the data of the CHOOZ experiment.

The $X^2$ of the solar neutrino rates,
$X^2_{\mathrm{S}}$,
is calculated using the procedure
described in
Refs.~\cite{Fogli-Lisi-correlations-95,%
Fogli-Lisi-Montanino-Palazzo-sun-3nu-00,%
Garzelli-Giunti-cs-00,%
Garzelli-Giunti-sf-00}:
\begin{equation}
X^2_{\mathrm{S}}
=
\sum_{j_1,j_2=1}^{N_{\mathrm{S}}}
\left( R^{\mathrm{(ex)}}_{j_1} - R^{\mathrm{(th)}}_{j_1} \right)
(V^{-1}_{\mathrm{S}})_{j_1j_2}
\left( R^{\mathrm{(ex)}}_{j_2} - R^{\mathrm{(th)}}_{j_2} \right)
\,,
\label{X2sun}
\end{equation}
where $V_{\mathrm{S}}$ is the covariance matrix of
experimental and theoretical uncertainties,
$R^{\mathrm{(ex)}}_{j}$
is the event rate measured in the $j^{\mathrm{th}}$ experiment
and
$R^{\mathrm{(th)}}_{j}$
is the corresponding theoretical event rate,
that depends on
$\Delta{m}^2$ and $\tan^2\theta$.
The indices $j,j_1,j_2=1,\ldots,N_{\mathrm{S}}$
with
$N_{\mathrm{S}}=4$
indicate the four types of solar neutrino experiments
listed in order in Table~\ref{rates}
together with the corresponding event rates
and experimental uncertainties.

The covariance matrix $V_{\mathrm{S}}$
takes into account the experimental and theoretical
uncertainties added in quadrature\footnote{
This is the common practice,
in spite of the impossibility to treat theoretical
uncertainties in the framework of Frequentist Statistics
(see item~\ref{6} in Section~\ref{Introduction}).
}
\cite{Fogli-Lisi-correlations-95,%
Fogli-Lisi-Montanino-Palazzo-sun-3nu-00,
Garzelli-Giunti-cs-00,%
Garzelli-Giunti-sf-00}:
\begin{eqnarray}
(V_{\mathrm{S}})_{j_1 j_2}
=
\null & \null & \null
\delta_{j_1 j_2}
\sigma^2_{j_1}
+
\delta_{j_1 j_2}
\left(
\sum_{i_1=1}^{8}
R_{i_1 j_1}^{\mathrm{(th)}}
\Delta\!\ln\!C_{i_1 j_1}^{\mathrm{(th)}}
\right)^2
\nonumber
\\
\null & \null & \null
+
\sum_{i_1,i_2=1}^{8}
R_{i_1 j_1}^{\mathrm{(th)}}
R_{i_2 j_2}^{\mathrm{(th)}}
\sum_{k=1}^{12}
\alpha_{i_1 k}
\alpha_{i_2 k}
\left( \Delta\!\ln\!X_k \right)^2
\,,
\label{VSrates}
\end{eqnarray}
where
$\sigma^2_{j}$
are the experimental uncertainties
given in Table~\ref{rates},
calculated by adding in quadrature the statistical and systematic
uncertainty for each experiment\footnote{
Also this common practice is completely unjustified
in the framework of Frequentist Statistics
(see item~\ref{6} in Section~\ref{Introduction}).
}.
The indices $i,i_1,i_2=1,\ldots,8$
denote the solar neutrino fluxes produced in the eight
solar thermonuclear reactions
$pp$,
$pep$,
$hep$,
$^7\mathrm{Be}$,
$^8\mathrm{B}$,
$^{13}\mathrm{N}$,
$^{15}\mathrm{O}$,
$^{17}\mathrm{F}$,
respectively.
The index $k=1,\ldots,12$
labels the twelve input astrophysical parameters $X_k$
in the SSM
($S_{1 1}$,
$S_{3 3}$,
$S_{3 4}$,
$S_{1\,14}$,
$S_{1 7}$,
Luminosity,
$Z/X$,
Age,
Opacity,
Diffusion,
$C_{^7\mathrm{Be}}$,
$S_{1\,16}$),
whose relative uncertainties
$\Delta\!\ln\!X_k$
determine the correlated uncertainties of the neutrino fluxes
$\Phi_{i}^{\mathrm{SSM}}$
through the logarithmic derivatives
\begin{equation}
\alpha_{ik}
=
\frac{\partial\ln\!\Phi_{i}^{\mathrm{SSM}}}{\partial\ln\!X_k}
\,.
\label{alpha}
\end{equation}
We adopt the values of
$\alpha_{ik}$
and
$\Delta\!\ln\!X_k$
given in Ref.~\cite{Fogli-Lisi-Montanino-Palazzo-sun-3nu-00},
except for $\Delta\!\ln\!X_7$ (relative to $Z/X$),
whose value has been updated in the BP2000 SSM \cite{BP2000}
from 0.033 to 0.061,
and
$\alpha_{i\,12}=\delta_{i8}$,
$\Delta\!\ln\!X_{12}=0.181$,
that have been introduced for the first time in
the BP2000 SSM \cite{BP2000}.
$X_{12}=S_{1\,16}$
is the $S$-factor for the reaction
$^{16}\mathrm{O}(p,\gamma)^{17}\mathrm{F}$
that determines the small $^{17}\mathrm{F}$ neutrino flux ($i=8$).
With these values we calculate the fractional uncertainties
of the SSM neutrino fluxes
\begin{equation}
\Delta\!\ln\!\Phi_{i}^{\mathrm{SSM}}
=
\sqrt{
\sum_{k=1}^{12}
\left( \alpha_{ik} \, \Delta\!\ln\!X_k \right)^2
}
\label{flux-uncertainties}
\end{equation}
listed in Table~\ref{uncertainties}.
One can see that these fractional uncertainties
are in good agreement
with those
presented in Table~7 of Ref.~\cite{BP2000}.
The uncertainty of the $hep$
flux is rather small,
in view of the fact that no uncertainty is given
in Ref.~\cite{BP2000}
because of the difficulty in calculating the $S$-factor
of the $hep$ reaction.
However,
this is irrelevant for our calculation
because in any case the contribution of the
$hep$ flux to solar neutrino data is negligible
(the Super-Ka\-mio\-kan\-de collaboration \cite{SK-sun-01} found that
the flux of $hep$ neutrinos on Earth
is less than 4.3 times the BP2000 SSM prediction at 90\% CL,
that is one hundred times smaller than the
$^8\mathrm{B}$ neutrino flux).

\TABULAR[t]{|c|c|c|c|c|c|c|c|c|}{
\hline
$i$&1&2&3&4&5&6&7&8
\\
&
$pp$
&
$pep$
&
$hep$
&
$^7\mathrm{Be}$
&
$^8\mathrm{B}$
&
$^{13}\mathrm{N}$
&
$^{15}\mathrm{O}$
&
$^{17}\mathrm{F}$
\\
\hline
$\Delta\!\ln\!\Phi_{i}^{\mathrm{SSM}}$
&
0.010
&
0.016
&
0.033
&
0.100
&
0.176
&
0.184
&
0.209
&
0.239
\vphantom{$\Big|$}
\\
\hline
}
{
\label{uncertainties}
Relative uncertainties
$\Delta\!\ln\!\Phi_{i}^{\mathrm{SSM}}$
of the neutrino fluxes obtained with
Eq.~(\protect\ref{flux-uncertainties}).
They are in good agreement
with those presented in Table~7 of Ref.~\protect\cite{BP2000}.
}

The quantity
\begin{equation}
R_{ij}^{\mathrm{(th)}}
=
\Phi_{i}^{\mathrm{SSM}}
\
C_{ij}^{\mathrm{(th)}}
\
P_{ij}^{\mathrm{(th)}}(\Delta{m}^2,\tan^2\theta)
\,,
\label{Rij}
\end{equation}
such that
\begin{equation}
R_{j}^{\mathrm{(th)}}
=
\sum_{i=1}^{8}
R_{ij}^{\mathrm{(th)}}
\,,
\label{Rj}
\end{equation}
is the theoretical event rate in the $j^{\mathrm{th}}$
experiment
due to the neutrino flux
$\Phi_{i}^{\mathrm{SSM}}$
produced in the $i^{\mathrm{th}}$
thermonuclear reaction in the sun
according to the BP2000 Standard Solar Model \cite{BP2000}.
$C_{ij}^{\mathrm{(th)}}$
is the corresponding energy-averaged cross section
and
$P_{ij}^{\mathrm{(th)}}(\Delta{m}^2,\tan^2\theta)$
is the corresponding averaged survival probability of solar $\nu_e$'s,
that depends on $\Delta{m}^2$ and $\tan^2\theta$
(in the case of the Super-Ka\-mio\-kan\-de experiment, $j=4$,
also the averaged $\nu_e\to\nu_a$ transition probability
must be properly taken into account).
The quantity
$
\Delta\ln\!C_{ij}^{\mathrm{(th)}}
=
\Delta\!C_{ij}^{\mathrm{(th)}} / C_{ij}^{\mathrm{(th)}}
$
is the relative uncertainty
of the energy-averaged cross section $C_{ij}^{\mathrm{(th)}}$.
The values of
$\Delta\ln\!C_{ij}^{\mathrm{(th)}}$
for $j=1,2,4$
(${}^{37}\mathrm{Cl}$,
${}^{71}\mathrm{Ga}$
and
$\mathrm{H}_2\mathrm{O}$
experiments)
are given in Ref.~\cite{Fogli-Lisi-Montanino-Palazzo-sun-3nu-00}.
For the SNO experiment
($j=3$)
$
\Delta\ln\!C_{33}^{\mathrm{(th)}}
=
\Delta\ln\!C_{53}^{\mathrm{(th)}}
=
3.0 \times 10^{-2}
$
\cite{SNO-01}
and the others are zero.

For the calculation of the probabilities
$P_{ij}^{\mathrm{(th)}}(\Delta{m}^2,\tan^2\theta)$
we have used the tables of neutrino spectra, solar 
density and radiochemical detector cross sections available in
Bahcall's web pages \cite{Bahcall-WWW},
BP2000 Standard Solar Model \cite{BP2000}.
For the calculation of the theoretical rate of the SNO experiment
we used the charged-current cross section
given in Refs.~\cite{Kubodera-deuteron-01,Kubodera-www}.
The probability of neutrino oscillations
is calculated with an unified approach that allows to
pass continuously from the vacuum oscillation regime to MSW transitions
\cite{Wolfenstein:1978ue,Mikheev:1985gs,Mikheev:1986wj} 
through the quasi-vacuum regime
\cite{Friedland-vo-00,Fogli-Lisi-Montanino-Palazzo-Quasi-vacuum-00},
using the quasi-vacuum analytical
prescription given in Ref.~\cite{Lisi:2000su}
(see also Refs.~\cite{Petcov:1988wv,Petcov:1989du}),
the usual prescription
for the MSW survival probability
(see
\cite{Fogli-Lisi-Montanino-Palazzo-Quasi-vacuum-00,%
Gonzalez-Garcia:2000sk})
and the level crossing probability 
appropriate for an exponential density profile
\cite{Petcov-analytic-87,Kuo-Pantaleone-RMP-89}.
We calculate the regeneration in the Earth
using a two-step model of the Earth density profile
\cite{Liu-Maris-Petcov-earth1-97,Petcov-diffractive-98,%
Akhmedov-parametric-99,%
Chizhov:1999az,Chizhov:1999he},
that is known to produce results that do not differ appreciably
from those obtained with
a less approximate
density profile.

The $X^2$ of the CHOOZ positron spectra,
$X^2_{\mathrm{C}}$
is calculated as in the analysis A of Ref.~\cite{Apollonio:1999ae},
with the following approximations.
Since we do not know the
antineutrino spectrum,
the spatial distribution functions of the reactor cores and detector
and the detector response function linking the real and visible
positron energies,
for each energy bin we calculated the oscillation
probability at the average energy
of the bin and the average distance of
the detector from each of the two reactors
\cite{Apollonio:1998xe}.
This approximation is quite good
because we are interested in values of
$\Delta{m}^2$ below $10^{-3} \, \mathrm{eV}$,
for which the energy and distance dependence of
the survival probability of the $\bar\nu_e$'s
in the CHOOZ experiment is very weak.
We calculate $X^2_{\mathrm{C}}$
as in Eq.~(13) of Ref.~\cite{Apollonio:1999ae},
with the only difference that we neglect the energy-scale
calibration factor,
whose small uncertainty (1.1\%) is practically negligible:
\begin{equation}
X^2_{\mathrm{C}}
=
\sum_{j_1,j_2=1}^{N_{\mathrm{C}}}
\left(
R^{\mathrm{(th)}}_{j_1}
-
\alpha_{\mathrm{C}}
\,
R^{\mathrm{(ex)}}_{j_1}
\right)
(V^{-1}_{\mathrm{C}})_{j_1j_2}
\left(
R^{\mathrm{(th)}}_{j_2}
-
\alpha_{\mathrm{C}}
\,
R^{\mathrm{(ex)}}_{j_2}
\right)
+
\left(
\frac{\alpha_{\mathrm{C}}-1}{\sigma_{\alpha_{\mathrm{C}}}}
\right)^2
\,,
\label{X2CHOOZ}
\end{equation}
where
$N_{\mathrm{C}} = 14$
is the number of energy bins
and
$\alpha_{\mathrm{C}}$
is the absolute normalization constant
with uncertainty
$\sigma_{\alpha_{\mathrm{C}}} = 2.7 \times 10^{-2}$
\cite{Apollonio:1999ae}.
We calculate the CHOOZ covariance matrix
$V_{\mathrm{C}}$
as described in Eq.~(12) of Ref.~\cite{Apollonio:1999ae}.
The only missing information in Ref.~\cite{Apollonio:1999ae}
is the value of the systematic uncertainties of the positron energy bins,
for which only the values for positron energy
2 and 6 MeV are given.
For the other bins we take systematic uncertainties
interpolated linearly between these two values.

The values of the minimum
$X^2_{\mathrm{min}}$
of the least-squares function (\ref{X2})
and the corresponding best-fit values of the oscillation parameters
for $\nu_e\to\nu_a$
or
$\nu_e\to\nu_s$
transitions are given in the first two rows of Table~\ref{chi2_min}.
The contribution to $X^2_{\mathrm{min}}$
from the CHOOZ least-squares function
$X^2_{\mathrm{C}}$
is always 6.0,
in reasonable agreement
with the minimum value
$(X^2_{\mathrm{min}})_{\mathrm{CHOOZ}}=5.0$
and
$(X^2_{\mathrm{no-oscillations}})_{\mathrm{CHOOZ}}=5.5$
found by the CHOOZ collaboration in Ref.~\cite{Apollonio:1999ae}.

\TABULAR[t]{|c|c|c|c|c|c|c|c|}{
\hline
\vphantom{$\Big|$}
Analysis
&
$X^2_{\mathrm{min}}$
&
$\tan^2\!\vartheta$
&
$\Delta{m}^2$ $(\mathrm{eV}^2)$
&
n.d.f.
&
g.o.f.
\\
\hline
\vphantom{$\Big|$}
\begin{tabular}{c}
Active
\\
Rates
\end{tabular}
&
$8.8$
&
$0.31 , 3.2$
&
$7.7 \times 10^{-11}$
&
$15$
&
$89\%$
\\
\hline
\vphantom{$\Big|$}
\begin{tabular}{c}
Sterile
\\
Rates
\end{tabular}
&
$11.3$
&
$0.36 , 2.8$
&
$1.1 \times 10^{-10}$
&
$15$
&
$73\%$
\\
\hline
\vphantom{$\Big|$}
\begin{tabular}{c}
Active
\\
Global
\end{tabular}
&
$38.4$
&
$0.38$
&
$6.3 \times 10^{-5}$
&
$52$
&
$92\%$
\\
\hline
\vphantom{$\Big|$}
\begin{tabular}{c}
Sterile
\\
Global
\end{tabular}
&
$52.8$
&
$0.36,2.8$
&
$4.9 \times 10^{-10}$
&
$52$
&
$44\%$
\\
\hline
}
{
\label{chi2_min}
Minima $X^2_{\mathrm{min}}$ of the least-squares function $X^2$
for $\nu_e\to\nu_a$ (Active)
or
$\nu_e\to\nu_s$ (Sterile)
in the Rates Analysis of Section~\ref{chi2-rates}
and in the Global Analysis of Section~\ref{chi2-global}.
We give also the number of degrees of freedom (n.d.f.),
and the corresponding goodness of fit (g.o.f.)
assuming for $X^2_{\mathrm{min}}$
a $\chi^2$ distribution with n.d.f. degrees of freedom.
}

One can see from Table~\ref{chi2_min}
that the minimum $X^2_{\mathrm{min}}$
lies in the Vacuum Oscillation (VO) region
for both Active and Sterile transitions.
Therefore,
there are two minima corresponding to the same value of $\sin^22\vartheta$
(they are connected by
$\log\tan^2\!\vartheta \to -\log\tan^2\!\vartheta$).

As written in Table~\ref{chi2_min},
the number of degrees of freedom (n.d.f.) in the Rates Analysis
is 15,
given by 4 solar rates plus 14 CHOOZ bins minus
3 parameters
($\alpha_{\mathrm{C}}$, $\tan^2\!\vartheta$, $\Delta{m}^2$).
The goodness of fit obtained
assuming for $X^2_{\mathrm{min}}$
a $\chi^2$ distribution with n.d.f. degrees of freedom
(see Refs.~\cite{Garzelli-Giunti-sf-00,Garzelli-Giunti-ugs-01}
for a discussion on the approximate character of this assumption)
is more than acceptable for both
Active and Sterile transitions.

However,
the large value of the goodness of fit
in the first two rows in Table~\ref{chi2_min}
looks suspicious:
if the assumption of a $\chi^2$ distribution with n.d.f. degrees of freedom
for
$X^2_{\mathrm{min}}$
is approximately correct,
the expected value of $X^2_{\mathrm{min}}$
is equal to $\mathrm{n.d.f.} = 15$,
which would give a goodness of fit of about 45\%.
The small value of $X^2_{\mathrm{min}}$
is mainly due to the inclusion in the fit of the
14 energy bins of the CHOOZ experiment,
that contribute with 13 degrees of freedom
and give a contribution of only 6.0 to $X^2_{\mathrm{min}}$.
This problem is already present in the analysis
of CHOOZ data performed by the CHOOZ collaboration
\cite{Apollonio:1999ae},
where a value of $(X^2_{\mathrm{min}})_{\mathrm{CHOOZ}}=5.0$
was found with 11 degrees of freedom.

From the second row in Table~\ref{chi2_min}
and taking into account
that the fit of the four solar neutrino rates
contributes with 5.3 to the value of
$X^2_{\mathrm{min}}$,
one could conclude that the explanation of the solar neutrino problem
in terms of $\nu_e\to\nu_s$ transitions
is acceptable.
This conclusion would be
in contradiction with the model-independent
evidence in favor of the presence of
active $\nu_{\mu,\tau}$
in the flux of solar neutrinos on Earth
discussed in Refs.~\cite{SNO-01,Fogli:2001vr,Giunti-aoe-01}.
This apparent contradiction is resolved by noting that
the model-independent calculations
performed in Refs.~\cite{SNO-01,Fogli:2001vr,Giunti-aoe-01}
are based on the adjustment of the Super-Ka\-mio\-kan\-de
electron energy threshold in order to make the
response functions of the SNO and Super-Ka\-mio\-kan\-de experiments
to solar neutrinos approximately equal
\cite{Villante:1998pe,Fogli:2001nn}.
The authors of Ref.~\cite{Fogli:2001vr} found that,
given the SNO threshold
$T_e^{\mathrm{SNO}} = 6.75 \, \mathrm{MeV}$,
the SNO and Super-Ka\-mio\-kan\-de response functions are approximately equal for
the Super-Ka\-mio\-kan\-de threshold
$T_e^{\mathrm{SK}} = 8.6 \, \mathrm{MeV}$,
instead of
$T_e^{\mathrm{SK}} = 4.5 \, \mathrm{MeV}$
for which the rate in the fourth row of Table~\ref{rates}
has been obtained \cite{SK-sun-01}.
The low $X^2_{\mathrm{min}}$
that we find in the VO region
is due to the strong energy dependence of the
$\nu_e\to\nu_s$
transition probability in the case of vacuum oscillations,
such that the averaged transition probability
is quite different in the SNO and Super-Ka\-mio\-kan\-de
experiments if their response function are different.
Indeed,
for the values of
$\tan^2\!\vartheta$
and
$\Delta{m}^2$
corresponding to
$X^2_{\mathrm{min}}$
we have
$\langle P_{\nu_e\to\nu_e} \rangle_{\mathrm{SNO}} = 0.347$
and
$\langle P_{\nu_e\to\nu_e} \rangle_{\mathrm{SK}} = 0.504$.

We have repeated our least-squares analysis
using $T_e^{\mathrm{SK}} = 8.6 \, \mathrm{MeV}$
and the corresponding rate
$0.451 \pm 0.017$
\cite{SK-sun-01,Fogli:2001vr}
for the Super-Ka\-mio\-kan\-de experiment,
normalized to the BP2000 SSM prediction,
and we found
$X^2_{\mathrm{min}} = 14.7$
for
$\tan^2\!\vartheta = 1.4 \times 10^{-3}$
and
$\Delta{m}^2 = 6.0 \times 10^{-6} \, \mathrm{eV^2}$,
\textit{i.e.} in the SMA region.
Now the contribution to $X^2_{\mathrm{min}}$
from the fit of the four solar neutrino rates
is 8.7,
in reasonable agreement with the model-independent
$3.06\sigma$
exclusion of pure
$\nu_e\to\nu_s$
transitions discussed in
Refs.~\cite{SNO-01,Fogli:2001vr,Giunti-aoe-01},
which would imply a contribution of about 9.4 to $X^2_{\mathrm{min}}$.
The discrepancy is due to the small remaining
difference of the SNO and Super-Ka\-mio\-kan\-de
response functions
(see Fig.~1 of Ref.~\cite{Fogli:2001vr}),
such that the averaged $\nu_e\to\nu_s$ transition probabilities
in the SNO and Super-Ka\-mio\-kan\-de experiments
are not exactly equal:
for the values of
$\tan^2\!\vartheta$
and
$\Delta{m}^2$
corresponding to
$X^2_{\mathrm{min}}$
we have
$\langle P_{\nu_e\to\nu_e} \rangle_{\mathrm{SNO}} = 0.453$
and
$\langle P_{\nu_e\to\nu_e} \rangle_{\mathrm{SK}} = 0.459$.
Indeed,
using ``by hand'' the same averaged probabilities of
$\nu_e\to\nu_s$ transitions
for the SNO and Super-Ka\-mio\-kan\-de experiments,
we obtain
$X^2_{\mathrm{min}} = 15.5$,
corresponding to a contribution of 9.5
from the fit of the four solar neutrino rates.

Even
using ``by hand'' the same averaged probabilities of
$\nu_e\to\nu_s$ transitions
for the SNO and Super-Ka\-mio\-kan\-de experiments,
the goodness of fit is high, 42\%.
Therefore,
we see that in the case of
$\nu_e\to\nu_s$ transitions
the
goodness of fit test is misleading,
because it does not exclude pure
Sterile transitions
that are excluded at $3.06\sigma$
with a model-independent comparison of Super-Ka\-mio\-kan\-de and SNO data
\cite{SNO-01,Fogli:2001vr,Giunti-aoe-01}.

The goodness of fit test is considered to be appropriate
for testing an hypothesis against all possible alternatives
(see \cite{Eadie-71}).
However,
in the case of solar neutrino oscillations
considered here we have two alternative hypotheses:
Active $\nu_e\to\nu_a$
and
Sterile $\nu_e\to\nu_s$ transitions.
An appropriate test for alternative hypotheses is the likelihood ratio
test\footnote{
We note, in passing,
that in the case of Active transitions
the maximum likelihood is reached
in the LMA region
for
$\tan^2\!\vartheta = 0.30$
and
$\Delta{m}^2 = 1.7 \times 10^{-5} \, \mathrm{eV}^2$.
The difference of the location of the maximum likelihood
and the minimum of $X^2$
(see Table~\ref{chi2_min})
is due to the determinant
$|V_{\mathrm{S}}|$
in Eq.~(\ref{lik}),
that depends on
$\tan^2\!\vartheta$
and
$\Delta{m}^2$.
In view of the fact that using the maximum likelihood
the best fit of the oscillation parameters
in the Rates Analysis lies in the same region
as the best fit in the Global Analysis
(see Section~\ref{chi2-global}),
a calculation of the allowed regions
based on the likelihood \cite{Garzelli-Giunti-ugs-01}
may be more robust than the standard
calculation based on $X^2$.
We do not present it here
because we consider the Bayesian analysis
discussed in Section~\ref{Bayesian}
definitely superior.
},
for which we find
\begin{equation}
\frac{
\mathrm{Max}_{\tan^2\!\vartheta,\Delta{m}^2,\alpha_{\mathrm{C}}}
\,
p^{(\mathrm{S})}(\mathrm{D}|\tan^2\!\vartheta,\Delta{m}^2,\alpha_{\mathrm{C}})
}{
\mathrm{Max}_{\tan^2\!\vartheta,\Delta{m}^2,\alpha_{\mathrm{C}}}
\,
p^{(\mathrm{A})}(\mathrm{D}|\tan^2\!\vartheta,\Delta{m}^2,\alpha_{\mathrm{C}})
}
=
0.24
\,,
\label{lik-ratio-rates}
\end{equation}
with the probability distribution functions
\begin{equation}
p^{(\mathrm{T})}(\mathrm{D}|\tan^2\!\vartheta,\Delta{m}^2,\alpha_{\mathrm{C}})
=
\frac{
e^{-X^2_{\mathrm{S}}/2}
}{
(2\pi)^{N_{\mathrm{S}}/2}\sqrt{|V_{\mathrm{S}}|}
}
\,
\frac{
e^{-X^2_{\mathrm{C}}/2}
}{
(2\pi)^{N_{\mathrm{C}}/2}\sqrt{|V_{\mathrm{C}}|}
}
\,,
\label{lik}
\end{equation}
where
$\mathrm{D}$
represents the data,
and
$\mathrm{T}=\mathrm{A}$
for $\nu_e\to\nu_a$ transitions (Active)
and
$\mathrm{T}=\mathrm{S}$
for $\nu_e\to\nu_s$ transitions (Sterile).

In Eq.~(\ref{lik-ratio-rates})
we have considered Sterile transitions as the null hypothesis
and Active transitions as the alternative hypothesis,
in order to see if the null hypothesis should be rejected or not.
This choice is one of the arbitrary choices
necessary in the application of Frequentist Statistics,
whose justification is purely subjective.
For example,
one may reason in the framework of 2+2 four-neutrino schemes\footnote{
In order to compare Active and Sterile transitions
one must work in the framework of a scheme with four or more massive
neutrinos.
Three neutrinos ($\nu_e$, $\nu_\mu$, $\nu_\tau$) are not enough,
because the existence of sterile neutrinos would be excluded a priori.
}
(see
\cite{BGG-review-98,DGKK-99,Concha-foursolar-00,%
Giunti-Laveder-3+1-00,Gonzalez-Garcia:2001uy}
and references therein),
where transitions of active in sterile neutrinos is
predicted in solar or atmospheric neutrino experiments.
Since no such transitions have been seen in
atmospheric neutrino experiments
\cite{Fukuda:2000np,Ambrosio:2001je},
one could consider reasonable the null hypothesis
that the solar neutrino problem
is due to practically pure
$\nu_e\to\nu_s$ transitions\footnote{
A different point of view could be favored by other scientists,
as the authors of Ref.~\cite{Barger-Fate-2000},
where a very interesting 3+1 four neutrino scheme
with practically no transitions of active into sterile neutrinos in both
solar and atmospheric neutrino experiments has been proposed.
In general,
in the framework of 3+1 four neutrino schemes
there may be or not transitions of active into sterile neutrinos in
solar and atmospheric neutrino experiments.
Hence,
for solar neutrinos
one can consider arbitrarily
Active transitions as the null hypothesis
and Sterile transitions as the alternative hypothesis,
or vice versa.
}.

The interpretation of the value of the likelihood ratio
for hypothesis testing
is another subjective choice necessary in the application of
Frequentist Statistics,
since the likelihood does not represent a probability.
Usually one chooses an arbitrary value\footnote{ \label{asymptotic}
Sometimes the value of this small number can be justified
in the asymptotic limit of
a large number of observations
(see \cite{Eadie-71}).
Clearly this limit has absolutely no interest in the case
of solar neutrino experiments.
},
of the order of 0.1\% or 1\%,
below which the null hypothesis is rejected
in favor of the alternative hypothesis.
Clearly,
the large value of the likelihood ratio (\ref{lik-ratio-rates})
does not allow to reject the null hypothesis
of pure Sterile transitions.

However,
we expect better chances to reject the null hypothesis
of pure Sterile transitions
using for the Super-Ka\-mio\-kan\-de
experiment the energy threshold
$T_e^{\mathrm{SK}} = 8.6 \, \mathrm{MeV}$,
for which
the SNO and Super-Ka\-mio\-kan\-de response functions are approximately equal
\cite{Fogli:2001vr}.
Indeed,
we find
\begin{equation}
\frac{
\mathrm{Max}_{\tan^2\!\vartheta,\Delta{m}^2,\alpha_{\mathrm{C}}}
\,
p^{(\mathrm{S})}(\mathrm{D}(T_e^{\mathrm{SK}} = 8.6 \, \mathrm{MeV})|\tan^2\!\vartheta,\Delta{m}^2,\alpha_{\mathrm{C}})
}{
\mathrm{Max}_{\tan^2\!\vartheta,\Delta{m}^2,\alpha_{\mathrm{C}}}
\,
p^{(\mathrm{A})}(\mathrm{D}(T_e^{\mathrm{SK}} = 8.6 \, \mathrm{MeV})|\tan^2\!\vartheta,\Delta{m}^2,\alpha_{\mathrm{C}})
}
=
0.05
\,,
\label{lik-ratio-rates-86}
\end{equation}
that is significantly lower
than the value in Eq.~(\ref{lik-ratio-rates}),
but still not sufficient to reject the null hypothesis
of pure Sterile transitions.

Hence,
we see that in the Rates Analysis
of solar neutrino data in terms of neutrino oscillations
Frequentist methods
are not able to reject the hypothesis of pure Sterile transitions,
that are excluded at $3.06\sigma$
in a model-independent way
\cite{SNO-01,Fogli:2001vr,Giunti-aoe-01}.
On the other hand,
Bayesian Probability Theory
allows to assign probabilities to hypotheses
and,
as we will see in Section~\ref{Active_Or_Sterile},
allows to show clearly that
$\nu_e\to\nu_s$ transitions
are disfavored with respect to
$\nu_e\to\nu_a$ transitions.

Let us determine now the allowed values
for the oscillation parameters
$\tan^2\!\vartheta$ and $\Delta{m}^2$
in the case of $\nu_e\to\nu_a$ transitions.
As done in Refs.~\cite{Fogli:2001vr,%
Bandyopadhyay:2001aa,Creminelli:2001ij,Krastev:2001tv},
we do not consider $\nu_e\to\nu_s$ transitions
because they are excluded at $3.06\sigma$
in a model-independent way,
as discussed in Refs.~\cite{SNO-01,Fogli:2001vr,Giunti-aoe-01}.

In a standard least-squares analysis
the
$
100\beta\%
$
CL regions
in the
$\tan^2\!\vartheta$--$\Delta{m}^2$ plane
are given by the condition
\begin{equation}
X^2 \leq X^2_{\mathrm{min}} + \Delta{X^2}(\beta)
\,,
\label{001}
\end{equation}
where
$\beta$ is the Confidence Level (CL)
and
$\Delta{X^2}(\beta)$
is the value of $X^2$ such that
the cumulative $X^2$ distribution for 2
degrees of freedom
is equal to $\beta$:
$\Delta{X^2}(\beta) = 4.61, \, 5.99, \, 9.21, \, 11.83$
for
$
\beta
=
0.90 (1.64\sigma), \,
0.95 (1.96\sigma), \,
0.99 (2.58\sigma), \,
0.9973 (3\sigma)
$.

The allowed regions that we obtain with
this standard method in the case of
$\nu_e\to\nu_a$
transitions are shown in Fig.~\ref{acr-chi}.
One can see that,
although the minimum of $X^2$ lies in the VO region,
only small areas in the VO region
are allowed,
because the oscillation probability
is very sensitive to variations of
$\Delta{m}^2$ and $\tan^2\!\vartheta$.
The allowed Large Mixing Angle (LMA) region
is much larger,
because the sensitivity of the oscillation probability
to variations of
$\Delta{m}^2$ and $\tan^2\!\vartheta$
is weak
and the value of $X^2$ is low
($X^2 = 9.1$ in the local minimum in the LMA region,
for
$\tan^2\!\vartheta = 0.34$
and
$\Delta{m}^2 = 2.0 \times 10^{-5} \, \mathrm{eV}^2$).
Also the Small Mixing Angle (SMA) and LOW regions
are relatively large for the same reason
($X^2 = 10.2$ in the local minimum in the SMA region,
for
$\tan^2\!\vartheta = 1.4 \times 10^{-3}$
and
$\Delta{m}^2 = 8.1 \times 10^{-6} \, \mathrm{eV}^2$,
and
$X^2 = 12.8$ in the local minimum in the LOW region,
for
$\tan^2\!\vartheta = 0.60$
and
$\Delta{m}^2 = 1.4 \times 10^{-7} \, \mathrm{eV}^2$).
The main effect of the inclusion in the analysis
of the CHOOZ data is to limit
the LMA region at large CL below
$\Delta{m}^2 \sim 10^{-3} \, \mathrm{eV}^2$
(in Fig.~\ref{acr-chi},
as well as in the following
Figs.~\ref{acg-chi}, \ref{acr-yab}, \ref{acg-yab}).

\subsection{Global Analysis}
\label{chi2-global}

In our Global Analysis,
instead of the total rate of the Super-Ka\-mio\-kan\-de experiment
we consider the data on the
Super-Ka\-mio\-kan\-de day and night electron energy spectra
presented in Ref.~\cite{SK-sun-hep-ex-0103032},
that contain information on the total rate
plus the shape of the energy spectrum.
The least-squares function is written as in Eq.~(\ref{X2}),
with the CHOOZ contribution
$X^2_{\mathrm{C}}$
given in Eq.~(\ref{X2CHOOZ}).
The solar contribution
$X^2_{\mathrm{S}}$
is written as in Eq.~(\ref{X2sun}),
but now
the indexes $j_1$, $j_2$
run from 1 to $N_{\mathrm{S}}=41$.
The indexes $j_1, j_2 = 1, 2, 3$
refer to the first three rates in Table~\ref{rates}.
The indexes $j_1, j_2 = 4, \ldots, 22$
and the indexes $j_1, j_2 = 23, \ldots, 41$
refer, respectively, to the data on the
Super-Ka\-mio\-kan\-de day and night electron energy spectra
presented in Table~III of Ref.~\cite{SK-sun-hep-ex-0103032}.
The covariance matrix $V_{\mathrm{S}}$ is written as
\begin{eqnarray}
(V_{\mathrm{S}})_{j_1 j_2}
=
\null & \null & \null
\delta_{j_1 j_2}
\sigma^2_{j_1}
+
\delta_{j_1 j_2}
\left(
\sum_{i_1} R_{i_1 j_1}^{\mathrm{(th)}} \Delta\!\ln\!C_{i_1 j_1}^{\mathrm{(th)}}
\right)^2
\nonumber
\\
\null & \null & \null
+
\sum_{i_1,i_2}
R_{i_1 j_1}^{\mathrm{(th)}} R_{i_2 j_2}^{\mathrm{(th)}}
\sum_k
\alpha_{i_1 k}
\alpha_{i_2 k}
\left( \Delta\!\ln\!X_k \right)^2
\nonumber
\\
\null & \null & \null
+
\theta_{j_1-3}
\,
\theta_{j_2-3}
\left[
(\sigma^{\mathrm{(th)}}_{\mathrm{cor}})_{j_1}
\,
(\sigma^{\mathrm{(th)}}_{\mathrm{cor}})_{j_2}
+
(\sigma^{\mathrm{(sys)}}_{\mathrm{cor}})_{j_1}
\,
(\sigma^{\mathrm{(sys)}}_{\mathrm{cor}})_{j_2}
\right.
\nonumber
\\
\null & \null & \null
\left.
\hspace{2.5cm}
+
(\sigma^{\mathrm{(sys)}}_{\mathrm{unc}})_{j_1}^2
\left(
\delta_{j_1(j_2-19)}
+
\delta_{j_1(j_2+19)}
\right)
\right]
\,,
\label{VSglobal}
\end{eqnarray}
with
$\theta_j=1$ for $j>0$
and
$\theta_j=0$ otherwise.
The quantities for $j_1,j_2\leq3$
are as in Eq.~(\ref{VSrates}).
For $j=4,\ldots,41$
we have
$
\sigma^2_{j}
=
(\sigma^{\mathrm{(sta)}})_{j}^2
+
(\sigma^{\mathrm{(sys)}}_{\mathrm{unc}})_{j}^2
$.
The statistical uncertainties
$(\sigma^{\mathrm{(sta)}})_j$ (for $j=4,\ldots,41$)
are given
in the third and fourth
columns of Table~III in Ref.~\cite{SK-sun-hep-ex-0103032}.
The uncorrelated systematic uncertainties
$(\sigma^{\mathrm{(sys)}}_{\mathrm{unc}})_j$
(for $j=4,\ldots,41$)
are given by
$
R^{\mathrm{(ex)}}_{j}
\,
\delta_{j,\mathrm{unc}}
$,
with $R^{\mathrm{(ex)}}_{j}$
listed in the third and fourth columns
of Table~III in Ref.~\cite{SK-sun-hep-ex-0103032}
for $j=4,\ldots,22$ (day spectrum)
and
$j=23,\ldots,41$ (night spectrum).
We assumed that the
systematic uncertainties of the day and night bins
with the same energy are fully correlated.
The values of
$\delta_{j,\mathrm{unc}}$
are listed in the sixth column of Table~III in
Ref.~\cite{SK-sun-hep-ex-0103032}
(we took the bigger between the positive and negative values).

The correlated uncertainties\footnote{
We would like to thank
M.C. Gonzalez-Garcia and
C. Pena-Garay
for explaining to us the treatment of correlated uncertainties
in Ref.~\cite{Bahcall:2001zu}
and sharing with us their knowledge on the
systematic uncertainties published by the
Super-Ka\-mio\-kan\-de collaboration in Ref.~\cite{SK-sun-hep-ex-0103032},
that leaded to an improvement of our analysis
with respect to the first version of the paper
sent to the hep-ph electronic archive.
}
(for $j=4,\ldots,41$, corresponding to the Super-Ka\-mio\-kan\-de energy bins)
are divided in theoretical,
$(\sigma^{\mathrm{(th)}}_{\mathrm{cor}})_j$,
and experimental systematic,
$(\sigma^{\mathrm{(sys)}}_{\mathrm{cor}})_j$.
The theoretical correlated uncertainties
$
(\sigma^{\mathrm{(th)}}_{\mathrm{cor}})_j
=
R^{\mathrm{(th)}}_{j}
\,
\delta^{\mathrm{(th)}}_{j,\mathrm{cor}}
$
come from the calculation of the shape of the $^8\mathrm{B}$ energy spectrum
\cite{Bahcall:1996qv,Ortiz:2000nf}.
We list in Table~\ref{delta}
the values of
$\delta^{\mathrm{(th)}}_{j,\mathrm{cor}}$
that we have extracted
from Fig.~5 of Ref.~\cite{Ortiz:2000nf}.

\TABULAR[t]{|c|c|c|c|}{
\hline
$
\begin{array}{c}
j
\\
\mbox{day}
\end{array}
$
&
$
\begin{array}{c}
j
\\
\mbox{night}
\end{array}
$
&
$ \delta^{\mathrm{(th)}}_{j,\mathrm{cor}} $
&
$ \delta^{\mathrm{(ex)}}_{j,\mathrm{cor}} $
\\
\hline
 4 & 23 & 0.0039 & 0.0020 \\ \hline
 5 & 24 & 0.0039 & 0.0020 \\ \hline
 6 & 25 & 0.0047 & 0.0029 \\ \hline
 7 & 26 & 0.0055 & 0.0058 \\ \hline
 8 & 27 & 0.0062 & 0.0077 \\ \hline
 9 & 28 & 0.0070 & 0.0106 \\ \hline
10 & 29 & 0.0078 & 0.0134 \\ \hline
11 & 30 & 0.0086 & 0.0163 \\ \hline
12 & 31 & 0.0093 & 0.0203 \\ \hline
13 & 32 & 0.0101 & 0.0242 \\ \hline
14 & 33 & 0.0109 & 0.0292 \\ \hline
15 & 34 & 0.0117 & 0.0331 \\ \hline
16 & 35 & 0.0124 & 0.0381 \\ \hline
17 & 36 & 0.0132 & 0.0440 \\ \hline
18 & 37 & 0.0140 & 0.0500 \\ \hline
19 & 38 & 0.0148 & 0.0570 \\ \hline
20 & 39 & 0.0155 & 0.0640 \\ \hline
21 & 40 & 0.0163 & 0.0730 \\ \hline
22 & 41 & 0.0205 & 0.1058 \\ \hline
}
{
\label{delta}
Values of the relative theoretical uncertainties
$\delta^{\mathrm{(th)}}_{j,\mathrm{cor}}$
and experimental uncertainties
$\delta^{\mathrm{(ex)}}_{j,\mathrm{cor}}$
of the Super-Ka\-mio\-kan\-de energy bins
used in the Global Analysis
(see Section~\ref{chi2-global}).
}

The experimental correlated uncertainties are given by
$
(\sigma^{\mathrm{(sys)}}_{\mathrm{cor}})_j
=
R^{\mathrm{(ex)}}_{j}
\,
\delta^{\mathrm{(ex)}}_{j,\mathrm{cor}}
$.
Unfortunately,
the relative correlated systematic uncertainties
presented by the Super-Ka\-mio\-kan\-de collaboration
in the fifth column of Table~III in
Ref.~\cite{SK-sun-hep-ex-0103032}
contains the energy-dependent part of
the theoretical uncertainty
on the $^8\mathrm{B}$ neutrino spectrum given in Ref.~\cite{Ortiz:2000nf}.
In order to avoid double-counting,
we subtracted quadratically
the energy-dependent part of $\delta^{\mathrm{(th)}}_{j,\mathrm{cor}}$
(\textit{i.e.}
$
\delta^{\mathrm{(th)}}_{j,\mathrm{cor}}
-
\delta^{\mathrm{(th)}}_{4,\mathrm{cor}}
$,
where $j=4$ is the index of the first Super-Ka\-mio\-kan\-de energy bin)
from the relative correlated systematic uncertainties
in the fifth column of Table~III in
Ref.~\cite{SK-sun-hep-ex-0103032}
(for which we took the bigger between the positive and negative values).
The resulting values of
$\delta^{\mathrm{(ex)}}_{j,\mathrm{cor}}$
are listed in Table~\ref{delta}.

This procedure for the treatment of systematic correlated uncertainties
is the best one that we could find,
given the available information.
It differs in some way from the treatment adopted by other authors
\cite{Fogli:2001vr,Bahcall:2001zu,%
Bandyopadhyay:2001aa,Creminelli:2001ij,Krastev:2001tv}.
In any case we would like to remark that
the effects of these different treatments
are very small on the final results,
because the dominant source of systematic correlated uncertainties is due to
the theoretical calculation of the total $^8\mathrm{B}$ neutrino flux,
whose uncertainty is about 20\% \cite{BP2000}.
This uncertainty is taken into account in the third term
on the right-hand side of Eq.~(\ref{VSglobal})
(analogous to the last term
on the right-hand side of Eq.~(\ref{VSrates})).

The values of the minimum $X^2_{\mathrm{min}}$
of the least-squares function
in the Global Analysis
in terms of
$\nu_e\to\nu_a$
and
$\nu_e\to\nu_s$
oscillations
are given, respectively,
in the third and fourth row of Table~\ref{chi2_min}.
One can see that $X^2_{\mathrm{min}}$
lies in the LMA region in the case of Active transitions,
whereas it lies in the VO region in the case of
Sterile transitions.
Again we see that the standard goodness of fit test
is not able to exclude pure Sterile transitions.
The reason of the low values of $X^2_{\mathrm{min}}$
with respect to the number of degrees of freedom
is the high number of Super-Ka\-mio\-kan\-de
and CHOOZ energy bins whose data do not fluctuate enough.
This fact could be due to chance
or to an overestimation of systematic uncertainties.

However,
in the Global Analysis
the likelihood test allows to
reject the null hypothesis of
pure Sterile transitions in favor of the alternative hypothesis
of pure Active transitions.
Indeed, we find
\begin{equation}
\frac{
\mathrm{Max}_{\tan^2\!\vartheta,\Delta{m}^2,\alpha_{\mathrm{C}}}
\,
p^{(\mathrm{S})}(\mathrm{D}|\tan^2\!\vartheta,\Delta{m}^2,\alpha_{\mathrm{C}})
}{
\mathrm{Max}_{\tan^2\!\vartheta,\Delta{m}^2,\alpha_{\mathrm{C}}}
\,
p^{(\mathrm{A})}(\mathrm{D}|\tan^2\!\vartheta,\Delta{m}^2,\alpha_{\mathrm{C}})
}
=
4 \times 10^{-4}
\,,
\label{lik-ratio-global}
\end{equation}
that is small enough.

Notice that the high number of Super-Ka\-mio\-kan\-de
and CHOOZ energy bins is irrelevant for the performance
of the likelihood ratio test of alternative hypotheses.
In this sense,
the likelihood ratio test is more robust than
the goodness of fit test.

Assuming that pure Sterile transitions are exclud\-ed,
we present in Fig.~\ref{acg-chi}
the allowed regions in the $\tan^2\!\vartheta$--$\Delta{m}^2$
plane for Active transitions.
One can see that there is no allowed SMA region
and the allowed VO regions are very small.
The local minimum in the LOW region has
$X^2 = 42.1$,
for
$\tan^2\!\vartheta = 0.64$
and
$\Delta{m}^2 = 1.4 \times 10^{-7} \, \mathrm{eV}^2$,
and the local minimum in the VO region has
$X^2 = 43.9$,
for
$\tan^2\!\vartheta = 0.38, 2.6$
and
$\Delta{m}^2 = 4.9 \times 10^{-10} \, \mathrm{eV}^2$.

The authors of
Refs.~\cite{Fogli:2001vr,Bandyopadhyay:2001aa,Creminelli:2001ij}
did not find, as us,
any allowed SMA region,
whereas the authors of Refs.~\cite{Bahcall:2001zu,Krastev:2001tv}
found an allowed SMA region at 99.73\% CL.
These discrepancies are due to differences in the treatment of
the data and the experimental and theoretical uncertainties.
Of course,
we obtain an allowed SMA region at higher CL,
precisely at 99.98\% CL ($3.7\sigma$).
Hence it is clear that
the SMA solution of the solar neutrino problem
is strongly disfavored by present data.
We think that further discussions on the precise
confidence level at which the SMA region is excluded
are not significant in practice
and their conclusions can change drastically
with new data.

\section{Bayesian Analysis}
\label{Bayesian}

In this Section we present our results
on the fit of solar neutrino data
in the framework of Bayesian Probability Theory.
In Section~\ref{Active_Or_Sterile}
we compare 
the probabilities of
$\nu_e\to\nu_a$
and
$\nu_e\to\nu_s$
transitions,
and in Section~\ref{Credible}
we calculate the Bayesian allowed regions
(credible regions)
for the oscillation parameters
$\Delta{m}^2$ and $\tan^2\vartheta$
in the case of
$\nu_e\to\nu_a$
transitions.

Bayes Theorem allows to calculate
the \emph{posterior probability distribution function}
$
p(\mathrm{T}, \- \tan^2\!\vartheta, \- \Delta{m}^2 \- | \- \mathrm{D},
\- \mathrm{I})
$,
where
$\mathrm{T}$
indicates the type of transitions:
$\mathrm{T}=\mathrm{A}$
for $\nu_e\to\nu_a$ transitions (Active),
and
$\mathrm{T}=\mathrm{S}$
for $\nu_e\to\nu_s$ transitions (Sterile).
Here
$\mathrm{D}$
represents the data
and
$\mathrm{I}$
represent all the background information and assumptions
on solar physics, neutrino physics,
etc.
For our analysis Bayes Theorem can be written as
\begin{equation}
p(\mathrm{T},\tan^2\!\vartheta,\Delta{m}^2|\mathrm{D},\mathrm{I})
=
\frac{
p(\mathrm{D}|\mathrm{T},\tan^2\!\vartheta,\Delta{m}^2,\mathrm{I})
\,
p(\mathrm{T},\tan^2\!\vartheta,\Delta{m}^2|\mathrm{I})
}{
p(\mathrm{D}|\mathrm{I})
}
\,,
\label{Bayes1}
\end{equation}
where
$p(\mathrm{D}|\mathrm{T},\tan^2\!\vartheta,\Delta{m}^2,\mathrm{I})$
is the \emph{likelihood function}
and
$p(\mathrm{T},\tan^2\!\vartheta,\Delta{m}^2|\mathrm{I})$
is the \emph{prior probability distribution function}
(the inclusion of background information and assumptions
is necessary because
in Bayesian Probability Theory,
as in real life,
all probabilities are conditional).
The probability
$p(\mathrm{D}|\mathrm{I})$
is known as \emph{global likelihood}
and acts as a normalization constant.

Assuming a normal distribution of statistical and systematic errors,
the likelihood function
is given by
\begin{equation}
p(\mathrm{D}|\mathrm{T},\tan^2\!\vartheta,\Delta{m}^2,\mathrm{I})
=
\frac{
e^{-X^2_{\mathrm{S}}/2}
}{
(2\pi)^{N_{\mathrm{S}}/2}\sqrt{|V_{\mathrm{S}}|}
}
\,
\int \mathrm{d}\alpha_{\mathrm{C}}
\frac{
e^{-X^2_{\mathrm{C}}/2}
}{
(2\pi)^{N_{\mathrm{C}}/2}\sqrt{|V_{\mathrm{C}}|}
}
\,,
\label{sampling}
\end{equation}
where we have marginalized the nuisance parameter
$\alpha_{\mathrm{C}}$.
Here $N_{\mathrm{S}}$
is the number of solar data points
($N_{\mathrm{S}}=4$ in the Rates Analysis
and
$N_{\mathrm{S}}=41$ in the Global Analysis)
and
$N_{\mathrm{C}}=14$
is the number of CHOOZ data points.
$X^2_{\mathrm{S}}$
is the solar least-squares function
and $V_{\mathrm{S}}$ is the corresponding covariance matrix,
whose calculation in the Rates Analysis and in the Global Analysis
is explained, respectively,
in Sections~\ref{chi2-rates} and \ref{chi2-global}.
$X^2_{\mathrm{C}}$
is the CHOOZ least-squares function
and $V_{\mathrm{C}}$ is the corresponding covariance matrix,
whose calculation
is explained
in Section~\ref{chi2-rates}.

The prior probability distribution function
can be written as
\begin{equation}
p(\mathrm{T},\tan^2\!\vartheta,\Delta{m}^2|\mathrm{I})
=
p(\tan^2\!\vartheta,\Delta{m}^2|\mathrm{T},\mathrm{I})
\,
p(\mathrm{T}|\mathrm{I})
\,,
\label{prior}
\end{equation}
where
$p(\mathrm{T}|\mathrm{I})$
is the prior probability of
$\nu_e\to\nu_a$
($\mathrm{T}=\mathrm{A}$)
or
$\nu_e\to\nu_s$
transitions
($\mathrm{T}=\mathrm{S}$),
and
$p(\tan^2\!\vartheta,\Delta{m}^2|\mathrm{T},\mathrm{I})$
is the prior probability distribution function
of the parameters
$\tan^2\!\vartheta$ and $\Delta{m}^2$
given $\mathrm{T}$ and $\mathrm{I}$.

The prior probability distribution function
$p(\tan^2\!\vartheta,\Delta{m}^2|\mathrm{T},\mathrm{I})$
quantifies the prior knowledge
on the parameters
$\tan^2\!\vartheta$ and $\Delta{m}^2$.
Assuming neutrino mixing
(included in $\mathrm{I}$),
for both $\mathrm{T}=\mathrm{A}$ and $\mathrm{T}=\mathrm{S}$
we know that solar neutrino data are sensitive to
several different order of magnitude of
$\tan^2\!\vartheta$ and $\Delta{m}^2$,
through vacuum oscillations
for $\Delta{m}^2 \lesssim 10^{-8} \, \mathrm{eV}^2$
and large mixing angles
or resonant MSW transitions
for
$
10^{-8} \, \mathrm{eV}^2
\lesssim
\Delta{m}^2
\lesssim
10^{-3} \, \mathrm{eV}^2
$
and
$
10^{-4}
\lesssim
\tan^2\!\vartheta
\lesssim
10
$.
Therefore,
the most reasonable non-informative
prior probability distribution function\footnote{
We think that this is a prior on which the scientific community
can agree,
on the basis of the common knowledge of the mechanisms of
solar neutrino oscillations.
Indeed,
the authors of Ref.~\cite{Creminelli:2001ij}
have chosen, independently,
the same prior.
},
that we will use in the following,
is a flat distribution in the
$\log(\tan^2\!\vartheta)$--$\log(\Delta{m}^2)$ plane
for both $\mathrm{T}=\mathrm{A}$ and $\mathrm{T}=\mathrm{S}$.
In this case,
using Eq.~(\ref{prior}),
Eq.~(\ref{Bayes1})
becomes
\begin{equation}
p(\mathrm{T},\tan^2\!\vartheta,\Delta{m}^2|\mathrm{D},\mathrm{I})
=
\frac{
p(\mathrm{D}|\mathrm{T},\tan^2\!\vartheta,\Delta{m}^2,\mathrm{I})
\,
p(\mathrm{T}|\mathrm{I})
}{ \displaystyle
\sum_{\mathrm{T}=\mathrm{A},\mathrm{S}}
\int \mathrm{d}\!\log(\tan^2\!\vartheta) \, \mathrm{d}\!\log(\Delta{m}^2) \,
p(\mathrm{D}|\mathrm{T},\tan^2\!\vartheta,\Delta{m}^2,\mathrm{I})
\,
p(\mathrm{T}|\mathrm{I})
}
\,,
\label{Bayes2}
\end{equation}
where we have expressed the global likelihood
$p(\mathrm{D}|\mathrm{I})$
as the appropriate normalization constant
and all probabilities are calculated integrating
$p(\mathrm{T},\tan^2\!\vartheta,\Delta{m}^2|\mathrm{D},\mathrm{I})$
over
$\mathrm{d}\!\log(\tan^2\!\vartheta)$ $\mathrm{d}\!\log(\Delta{m}^2)$.

\subsection{Active Or Sterile?}
\label{Active_Or_Sterile}

In this Section we confront the probabilities
of Active
$\nu_e\to\nu_a$
and Sterile
$\nu_e\to\nu_s$
transitions
using the relation
\begin{equation}
p(\mathrm{T}|\mathrm{D},\mathrm{I})
=
\int \mathrm{d}\!\log(\tan^2\!\vartheta) \, \mathrm{d}\!\log(\Delta{m}^2) \,
p(\mathrm{T},\tan^2\!\vartheta,\Delta{m}^2|\mathrm{D},\mathrm{I})
\,.
\label{pT}
\end{equation}
From Eq.~(\ref{Bayes2}),
the ratio of the probabilities of
Sterile and Active transitions is given by
\begin{equation}
\frac
{ p(\mathrm{S}|\mathrm{D},\mathrm{I}) }
{ p(\mathrm{A}|\mathrm{D},\mathrm{I}) }
=
\frac
{
\int \mathrm{d}\!\log(\tan^2\!\vartheta) \, \mathrm{d}\!\log(\Delta{m}^2) \,
p(\mathrm{D}|\mathrm{S},\tan^2\!\vartheta,\Delta{m}^2,\mathrm{I})
}{
\int \mathrm{d}\!\log(\tan^2\!\vartheta) \, \mathrm{d}\!\log(\Delta{m}^2) \,
p(\mathrm{D}|\mathrm{A},\tan^2\!\vartheta,\Delta{m}^2,\mathrm{I})
}
\,
\frac
{ p(\mathrm{S}|\mathrm{I}) }
{ p(\mathrm{A}|\mathrm{I}) }
\,.
\label{ratio1}
\end{equation}
Notice that the ratio of the prior probabilities
of Sterile and Active transitions
factorizes out.
Since we do not have any prior preference for Sterile or Active transitions,
we take their prior probabilities as equal,
leading to
\begin{equation}
\frac
{ p(\mathrm{S}|\mathrm{D},\mathrm{I}) }
{ p(\mathrm{A}|\mathrm{D},\mathrm{I}) }
=
\frac
{
\int \mathrm{d}\!\log(\tan^2\!\vartheta) \, \mathrm{d}\!\log(\Delta{m}^2) \,
p(\mathrm{D}|\mathrm{S},\tan^2\!\vartheta,\Delta{m}^2,\mathrm{I})
}{
\int \mathrm{d}\!\log(\tan^2\!\vartheta) \, \mathrm{d}\!\log(\Delta{m}^2) \,
p(\mathrm{D}|\mathrm{A},\tan^2\!\vartheta,\Delta{m}^2,\mathrm{I})
}
\,.
\label{ratio}
\end{equation}

Our result in the Rates Analysis
($(\mathrm{D},\mathrm{I}) = \mathrm{Rates Analysis}$)
is
\begin{equation}
\frac
{ p(\mathrm{S}|\mathrm{Rates Analysis}) }
{ p(\mathrm{A}|\mathrm{Rates Analysis}) }
=
2.8 \times 10^{-2}
\,.
\label{ratio_aos_rates}
\end{equation}
It is clear that Sterile transitions
are disfavored with respect to Active transitions,
in agreement with the model-independent
exclusion at $3.06\sigma$ of pure Sterile transitions
\cite{SNO-01,Fogli:2001vr,Giunti-aoe-01}.
From the discussion in Section~\ref{chi2-rates},
it is clear that the incompatibility of
Sterile transitions with the data
should be enhanced by choosing for the Super-Ka\-mio\-kan\-de
experiment the energy threshold
$T_e^{\mathrm{SK}} = 8.6 \, \mathrm{MeV}$,
for which
the SNO and Super-Ka\-mio\-kan\-de response functions are approximately equal
\cite{Fogli:2001vr}.
Indeed, we find
\begin{equation}
\frac
{ p(\mathrm{S}|\mathrm{Rates Analysis},T_e^{\mathrm{SK}} = 8.6 \, \mathrm{MeV}) }
{ p(\mathrm{A}|\mathrm{Rates Analysis},T_e^{\mathrm{SK}} = 8.6 \, \mathrm{MeV}) }
=
1.5 \times 10^{-2}
\,.
\label{ratio_aos_rates_86}
\end{equation}
Notice that,
although the number in Eq.~(\ref{ratio_aos_rates_86})
is comparable with the one in the likelihood ratio
(\ref{lik-ratio-rates-86}),
the interpretation is very different.
Here we have a ratio of probabilities of hypotheses
and a very small ratio means that the probability
of Sterile transitions is much smaller than that
of Active transitions.
No such interpretation exist for the likelihood ratio
in the framework of Frequentist Statistics,
in which the null hypothesis can be either accepted or rejected
in favor of an alternative hypothesis
on the basis of the comparison of
the value of the likelihood ratio
with a very small fixed number
chosen arbitrarily in advance
(see footnote~\ref{asymptotic}).

Our result in the Global Analysis
($(\mathrm{D},\mathrm{I}) = \mathrm{Global Analysis}$)
is even stronger:
\begin{equation}
\frac
{ p(\mathrm{S}|\mathrm{Global Analysis}) }
{ p(\mathrm{A}|\mathrm{Global Analysis}) }
=
4 \times 10^{-4}
\,,
\label{ratio_aos_global}
\end{equation}
that practically excludes Sterile transitions
with respect to Active transitions.

Hence,
the Bayesian analysis of solar neutrino data shows that
Sterile transitions
are strongly disfavored with respect to Active transitions,
in agreement with the model-independent
exclusion at $3.06\sigma$ of pure Sterile transitions
\cite{SNO-01,Fogli:2001vr,Giunti-aoe-01}.
We see that Bayesian Probability Theory
gives an unambiguous and correct result
in the comparison of Active and Sterile transitions
in both the Rates and Global Analysis,
contrary to the popular goodness of fit test
used in traditional analyses of solar neutrino data
and the likelihood ratio test in the Rates Analysis,
as shown in Section~\ref{chi2}.

This better performance of Bayesian Probability Theory
in model comparison is first of all due to the fact that
in
Bayesian Probability Theory
as in real life
one can assign probabilities to
hypotheses and discuss which hypothesis
is more or less favored in comparison with others,
whereas in Frequentist Statistics
one is only allowed to either accept or reject an hypothesis.
Secondly,
the judgment of an hypothesis in the framework of
Bayesian Probability Theory
is more robust than in Frequentist Statistics
because Bayesian Probability Theory
allows to estimate the likelihood of a model
not only from the best-fit point of its parameter space
as in Frequentist methods,
but from the performance of the model averaged over
all its parameter space.
Thirdly,
in the framework of Bayesian Probability Theory
different hypotheses are fairly compared
assigning to them the same prior,
whereas in the likelihood ratio test
it is required to choose a null hypothesis and an alternative hypothesis,
that are treated quite differently.

Notice that
the Bayesian method
does not suffer any problem from the inclusion
of the numerous bins of the CHOOZ and Super-Ka\-mio\-kan\-de
electron energy spectra,
contrary to the least-squares method discussed in Section~\ref{chi2}.
\subsection{Credible Regions}
\label{Credible}

In this Section we present the results of our calculation
of Bayesian
\emph{credible regions}
(also known as
\emph{highest posterior density regions})
in the plane of the oscillation parameters
$\tan^2\!\vartheta$ and $\Delta{m}^2$
for Active transitions, that are strongly favored
over Sterile transitions,
as shown in Section~\ref{Active_Or_Sterile}.
The credible regions
are so-called in order to distinguish them
from the ``confidence regions''
calculated in the framework of Frequentist Statistics.
Credible regions contain a specified fraction of
the posterior probability and all values of
the parameters inside the credible regions
have higher probability than those outside.

The posterior probability distribution function for
the oscillation parameters
$\tan^2\!\vartheta$ and $\Delta{m}^2$
in the case of Active transitions
is given by
\begin{equation}
p(\tan^2\!\vartheta,\Delta{m}^2|\mathrm{A},\mathrm{D},\mathrm{I})
=
\frac
{ p(\mathrm{A},\tan^2\!\vartheta,\Delta{m}^2|\mathrm{D},\mathrm{I}) }
{ p(\mathrm{A}|\mathrm{D},\mathrm{I}) }
\,.
\label{posterior-parameters-Active-1}
\end{equation}
Assuming Active transitions implies that we take
the prior probability of Active transitions to be unity:
\begin{equation}
p(\mathrm{A}|\mathrm{I}) = 1
\,,
\label{prior-Active}
\end{equation}
that implies $p(\mathrm{S}|\mathrm{I}) = 0$.
Therefore,
we have
\begin{equation}
p(\mathrm{D}|\mathrm{I})
=
p(\mathrm{D},\mathrm{A}|\mathrm{I})
+
p(\mathrm{D},\mathrm{S}|\mathrm{I})
=
p(\mathrm{D}|\mathrm{A},\mathrm{I})
\,
p(\mathrm{A}|\mathrm{I})
+
p(\mathrm{D}|\mathrm{S},\mathrm{I})
\,
p(\mathrm{S}|\mathrm{I})
=
p(\mathrm{D}|\mathrm{A},\mathrm{I})
\,,
\label{pDI}
\end{equation}
and
\begin{equation}
p(\mathrm{A}|\mathrm{D},\mathrm{I})
=
\frac{
p(\mathrm{D}|\mathrm{A},\mathrm{I})
\,
p(\mathrm{A}|\mathrm{I})
}{
p(\mathrm{D}|\mathrm{I})
}
=
1
\,.
\label{posterior-Active}
\end{equation}
From Eqs.~(\ref{Bayes2}), (\ref{posterior-parameters-Active-1}),
(\ref{prior-Active}) and (\ref{posterior-Active})
we obtain
\begin{equation}
p(\tan^2\!\vartheta,\Delta{m}^2|\mathrm{A},\mathrm{D},\mathrm{I})
=
\frac{
p(\mathrm{D}|\mathrm{A},\tan^2\!\vartheta,\Delta{m}^2,\mathrm{I})
}{ \displaystyle
\int \mathrm{d}\!\log(\tan^2\!\vartheta) \, \mathrm{d}\!\log(\Delta{m}^2) \,
p(\mathrm{D}|\mathrm{A},\tan^2\!\vartheta,\Delta{m}^2,\mathrm{I})
}
\,.
\label{posterior-parameters-Active-2}
\end{equation}

Using the expression (\ref{sampling})
with
$\mathrm{T} = \mathrm{A}$
for
$p(\mathrm{D}|\mathrm{A},\tan^2\!\vartheta,\Delta{m}^2,\mathrm{I})$,
we obtained the credible regions
with
90\%,
95\%,
99\% and
99.73\%
posterior probability
shown in Fig.~\ref{acr-yab} for the Rates Analysis
and in Fig.~\ref{acg-yab} for the Global Analysis.
One can see that these regions are similar but larger
than the corresponding allowed regions
obtained with the standard least-squares analysis
presented in Figs.~\ref{acr-chi} and \ref{acg-chi}.
Let us emphasize,
however,
that the meaning of Bayesian and Frequentist regions
is quite different and a direct comparison is meaningless.

Bayesian Probability Theory
allows to calculate the relative probabilities of the
SMA, LMA, LOW and VO regions.
The posterior probability that
the true values of the oscillation parameters
lie in a region $\mathrm{R}$,
with
$\mathrm{R} = \mathrm{SMA}, \mathrm{LMA}, \mathrm{LOW}, \mathrm{VO}$,
is given by
\begin{equation}
p(\mathrm{R}|\mathrm{A},\mathrm{D},\mathrm{I})
=
\int_{\mathrm{R}}
\mathrm{d}\!\log(\tan^2\!\vartheta) \, \mathrm{d}\!\log(\Delta{m}^2) \,
p(\tan^2\!\vartheta,\Delta{m}^2|\mathrm{A},\mathrm{D},\mathrm{I})
\,,
\label{pR}
\end{equation}
where the integration is performed over the appropriate
ranges of the parameters given in Eqs.~(\ref{SMA})--(\ref{VO}).

In the Rates Analysis we found
\begin{eqnarray}
&
p(\mathrm{LMA}|\mathrm{A},\mathrm{D},\mathrm{I})
=
0.72
\,,
\nonumber
\\
&
p(\mathrm{VO}|\mathrm{A},\mathrm{D},\mathrm{I})
=
0.12
\,,
\nonumber
\\
&
p(\mathrm{SMA}|\mathrm{A},\mathrm{D},\mathrm{I})
=
0.10
\,,
\nonumber
\\
&
p(\mathrm{LOW}|\mathrm{A},\mathrm{D},\mathrm{I})
=
0.06
\,,
\label{pR-rates}
\end{eqnarray}
and in the Global Analysis
\begin{eqnarray}
&
p(\mathrm{LMA}|\mathrm{A},\mathrm{D},\mathrm{I})
=
0.86
\,,
\nonumber
\\
&
p(\mathrm{LOW}|\mathrm{A},\mathrm{D},\mathrm{I})
=
0.13
\,,
\nonumber
\\
&
p(\mathrm{VO}|\mathrm{A},\mathrm{D},\mathrm{I})
=
0.01
\,,
\nonumber
\\
&
p(\mathrm{SMA}|\mathrm{A},\mathrm{D},\mathrm{I})
=
10^{-4}
\,.
\label{pR-global}
\end{eqnarray}
Hence,
the LMA region is favored in the Global Analysis
as well as in the Rates Analysis
(see Ref.~\cite{Berezinsky-Lissia-01}
for an explanation of the physical reason).
This consistency in the Bayesian analysis of solar neutrino data
is very interesting and promising
for future experiments that
plan to explore the LMA region
with terrestrial experiments
\cite{KAMLAND,BOREXINO}.

\subsection{Marginal Bayesian Distributions}
\label{Marginal}

In this Section we marginalize the posterior probability
(\ref{posterior-parameters-Active-2})
in order to derive the separate posterior probability distributions
for
$\tan^2\!\vartheta$
and
$\Delta{m}^2$:
\begin{eqnarray}
\null & \null & \null
p(\tan^2\!\vartheta|\mathrm{A},\mathrm{D},\mathrm{I})
=
\int \mathrm{d}\!\log(\Delta{m}^2) \,
p(\tan^2\!\vartheta,\Delta{m}^2|\mathrm{A},\mathrm{D},\mathrm{I})
\,,
\label{margial1}
\\
\null & \null & \null
p(\Delta{m}^2|\mathrm{A},\mathrm{D},\mathrm{I})
=
\int \mathrm{d}\!\log(\tan^2\!\vartheta) \,
p(\tan^2\!\vartheta,\Delta{m}^2|\mathrm{A},\mathrm{D},\mathrm{I})
\,.
\label{marginal2}
\end{eqnarray}
The posterior probability distributions
that we obtained in the Rates Analysis are
shown in Figs.~\ref{acr-yab-x} and \ref{acr-yab-y}
and those that we obtained in the Global Analysis are
shown in Figs.~\ref{acg-yab-x} and \ref{acg-yab-y}.

From Fig.~\ref{acr-yab-x} one can see that
in the Rates Analysis there are two peaks of
the posterior probability distribution for
$\tan^2\!\vartheta$:
one at small mixing angles,
$\tan^2\!\vartheta \sim 10^{-3}$,
and one for large mixing angles,
$\tan^2\!\vartheta \sim 0.3$.
However,
large mixing angles are strongly favored,
as one can see calculating the corresponding probability:
\begin{equation}
p(0.1<\tan^2\!\vartheta<10|\mathrm{A},\mathrm{D},\mathrm{I})
=
0.90
\,.
\label{pt2t}
\end{equation}
In the case of the Global Analysis,
Fig.~\ref{acg-yab-x}
shows that there is only
one peak for
the posterior probability distribution for
$\tan^2\!\vartheta$
at large mixing angles,
with probability close to unity.

Figure~\ref{acr-yab-y} shows that in the Rates Analysis
the posterior probability distribution of
$\Delta{m}^2$
has a large and high peak for
$
2 \times 10^{-6} \, \mathrm{eV}^2
\lesssim
\Delta{m}^2
\lesssim
10^{-3} \, \mathrm{eV}^2
$,
a small and low peak for
$
\Delta{m}^2
\sim
10^{-7} \, \mathrm{eV}^2
$,
and several sharp peaks for
$
\Delta{m}^2
\lesssim
10^{-8} \, \mathrm{eV}^2
$
(due to the rapid oscillations of the transition
probability in vacuum as a function of
$\Delta{m}^2$).
A calculation of the corresponding probabilities
shows that large values of
$\Delta{m}^2$
are favored:
\begin{eqnarray}
\null & \null & \null
p(
2 \times 10^{-6} \, \mathrm{eV}^2
<
\Delta{m}^2
<
10^{-3} \, \mathrm{eV}^2
|\mathrm{A},\mathrm{D},\mathrm{I})
=
0.82
\,,
\nonumber
\\
\null & \null & \null
p(
10^{-8} \, \mathrm{eV}^2
<
\Delta{m}^2
<
2 \times 10^{-6} \, \mathrm{eV}^2
|\mathrm{A},\mathrm{D},\mathrm{I})
=
0.06
\,,
\nonumber
\\
\null & \null & \null
p(
\Delta{m}^2
<
10^{-8} \, \mathrm{eV}^2
|\mathrm{A},\mathrm{D},\mathrm{I})
=
0.12
\,.
\label{pdm2-rates}
\end{eqnarray}
From Fig.~\ref{acg-yab-y} one can see that
in the Global Analysis
large values of $\Delta{m}^2$
are even more favored than in the Rates Analysis
and very small values are strongly disfavored.
Indeed,
we obtained the probabilities
\begin{eqnarray}
\null & \null & \null
p(
2 \times 10^{-6} \, \mathrm{eV}^2
<
\Delta{m}^2
<
10^{-3} \, \mathrm{eV}^2
|\mathrm{A},\mathrm{D},\mathrm{I})
=
0.86
\,,
\nonumber
\\
\null & \null & \null
p(
10^{-8} \, \mathrm{eV}^2
<
\Delta{m}^2
<
2 \times 10^{-6} \, \mathrm{eV}^2
|\mathrm{A},\mathrm{D},\mathrm{I})
=
0.13
\,,
\nonumber
\\
\null & \null & \null
p(
\Delta{m}^2
<
10^{-8} \, \mathrm{eV}^2
|\mathrm{A},\mathrm{D},\mathrm{I})
=
0.01
\,.
\label{pdm2-global}
\end{eqnarray}

\section{Conclusions}
\label{Conclusions}

In this paper we have presented the results of a
Bayesian analysis
of solar neutrino data,
including the recently published rate of CC interactions
in the SNO experiment \cite{SNO-01},
and
the data of the CHOOZ experiment \cite{Apollonio:1999ae},
that exclude large values of the mixing angle
for $\Delta{m}^2 \gtrsim 10^{-3} \, \mathrm{eV}^2$.
We have considered
$\nu_e\to\nu_a$
oscillations,
with $a=\mu,\tau$,
and
$\nu_e\to\nu_s$
oscillations,
where $\nu_s$ is a sterile neutrino.
We have shown in Section~\ref{Active_Or_Sterile}
that the Bayesian analysis
implies that $\nu_e\to\nu_s$ transitions
are strongly disfavored with respect to
$\nu_e\to\nu_a$ transitions,
in agreement with the model-independent
$3.06\sigma$
exclusion of pure
$\nu_e\to\nu_s$
transitions discussed in
Refs.~\cite{SNO-01,Fogli:2001vr,Giunti-aoe-01}.

We have also presented,
in Section~\ref{chi2},
the results of a standard least-squares
analysis of solar neutrino data
and we have shown that the standard goodness of fit test
is not able to exclude pure
$\nu_e\to\nu_s$ transitions.
A problem of the goodness of fit test
is the small fluctuations
of the data relative to the numerous
bins of the CHOOZ and Super-Ka\-mio\-kan\-de
electron energy spectra.
The likelihood ratio test,
that is insensitive to the number of bins,
allows to reject the null hypothesis
of
pure
$\nu_e\to\nu_s$ transitions
in favor of
$\nu_e\to\nu_a$ transitions
only in the Global Analysis.

In Section~\ref{Credible}
we have presented
the results of our calculation
of Bayesian
credible regions
in the plane of the oscillation parameters
$\tan^2\!\vartheta$ and $\Delta{m}^2$
for $\nu_e\to\nu_a$ transitions
(Figs.~\ref{acr-yab} and \ref{acg-yab}).
The Bayesian credible regions are significantly larger
than the corresponding least-squares allowed region
(Figs.~\ref{acr-chi} and \ref{acg-chi}),
presented in Sections~\ref{chi2-rates} and \ref{chi2-global}.
However,
it is always necessary to keep in mind that a direct comparison of
Bayesian and Frequentist regions is meaningless,
because they have different properties
due to the different definitions of probability
in Bayesian Probability Theory and Frequentist Statistics.

Bayesian Probability Theory allows to calculate the probability
of separate regions in the plane of the oscillation parameters
$\tan^2\!\vartheta$ and $\Delta{m}^2$.
We found in the Global Analysis that the Large Mixing Angle (LMA)
region is strongly favored by the data
(86\% probability),
the LOW region has some small chance
(13\% probability),
the Vacuum Oscillation (VO) region
is almost excluded
(1\% probability)
and
the Small Mixing Angle (SMA) region is practically excluded
(0.01\% probability).
Also the Rates Analysis favors the LMA region.

We have also presented in Section~\ref{Marginal} the marginal
posterior probability distributions
for
$\tan^2\!\vartheta$
and
$\Delta{m}^2$.
In the Global Analysis
the data imply large mixing almost with certainty
and large values of
$\Delta{m}^2$
are favored
($
2 \times 10^{-6} \, \mathrm{eV}^2
<
\Delta{m}^2
<
10^{-3} \, \mathrm{eV}^2
$
with 86\% probability).
Also in the Rates Analysis large mixing and
large values of $\Delta{m}^2$
are favored.

These indications in favor of
large mixing and large values for $\Delta{m}^2$
are very encouraging
for future terrestrial experiments
as
KAMLAND \cite{KAMLAND}
and
BOREXINO
\cite{KAMLAND,BOREXINO}
that have the possibility to explore this region of parameter space.

In conclusion,
we want to emphasize the better
performance
shown in this paper
of Bayesian Probability Theory
with respect to Frequentist Statistics
in the analysis of solar neutrino data.
In particular,
the Bayesian analysis of solar neutrino data is able to
disfavor clearly Sterile transitions with respect to Active ones,
does not suffer any problem from the inclusion in the analysis
of the numerous bins of the CHOOZ and Super-Ka\-mio\-kan\-de
electron energy spectra,
allows to
reach the same conclusion on the determination of the most favored
values of the oscillation parameters
in both the Rates and Global Analysis.

\acknowledgments

M.V.G. thanks A. Palazzo for precious suggestions,
G.L. Fogli and E. Lisi for interesting discussions during
the First Italian Astroparticle School (Otranto 11-16 June, 2001).
We would like to express our gratitude to
M.C. Gonzalez-Garcia and
C. Pena-Garay for useful remarks on the first version of this paper.

\listoftables
\listoffigures

\providecommand{\href}[2]{#2}\begingroup\raggedright\endgroup

\newpage

\FIGURE{
\includegraphics[bb=120 505 372 788, width=\textwidth]{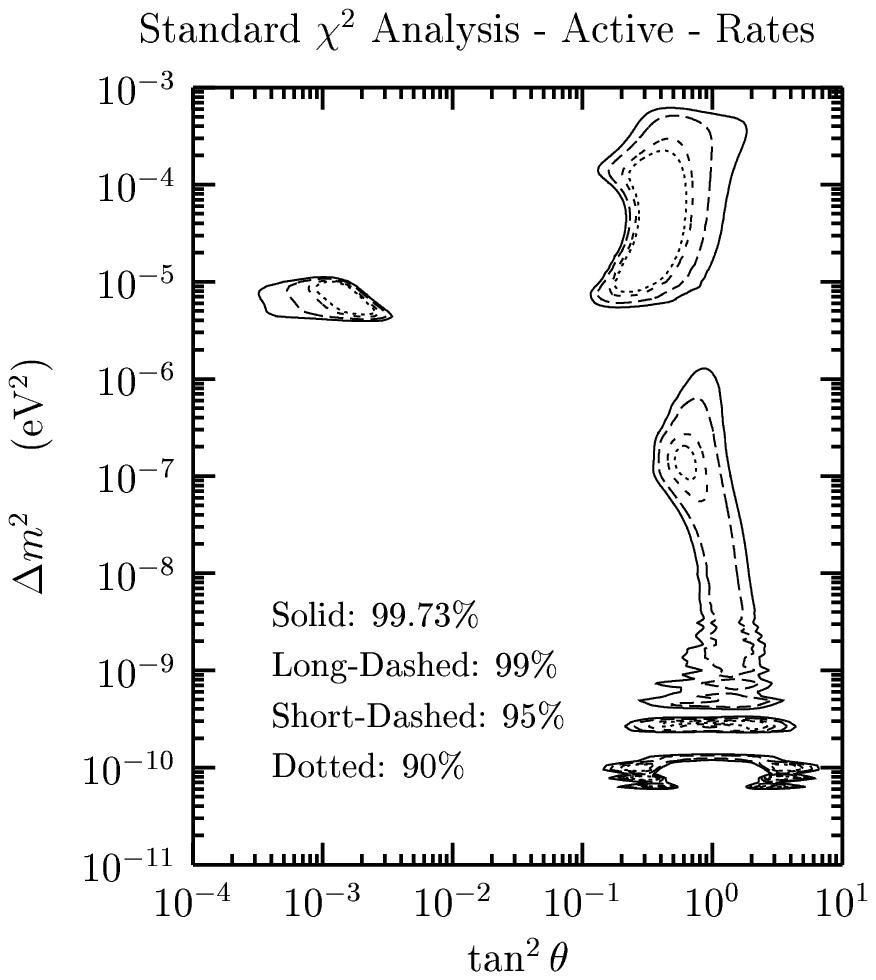}
\caption{
\label{acr-chi}
Allowed regions in the standard least-squares Rates Analysis
(Section~\ref{chi2-rates})
of solar neutrino rates and CHOOZ data
in terms of
$\nu_e\to\nu_a$
transitions.
}
}

\FIGURE{
\includegraphics[bb=120 505 372 788, width=\textwidth]{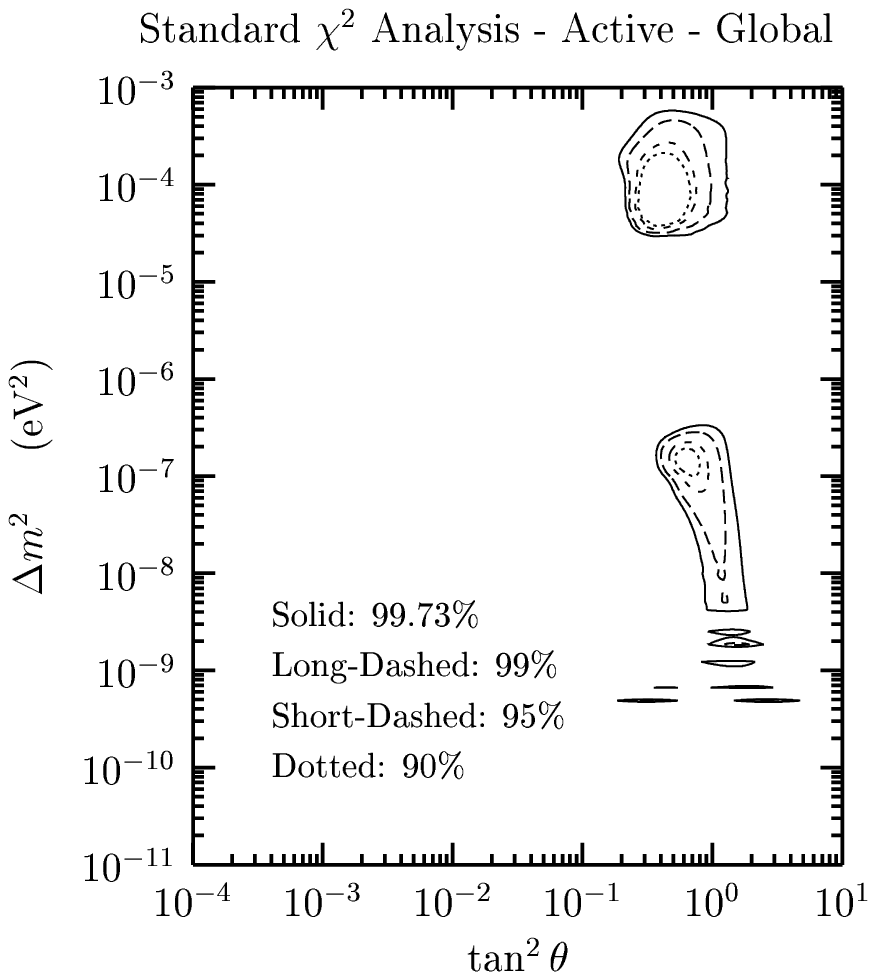}
\caption{
\label{acg-chi}
Allowed regions in the standard least-squares Global Analysis
(Section~\ref{chi2-global})
of the rates of the Homestake,
GALLEX+GNO+SAGE
and
SNO experiments,
the Super-Ka\-mio\-kan\-de day and night electron energy spectra
and CHOOZ data
in terms of
$\nu_e\to\nu_a$
transitions.
}
}

\FIGURE{
\includegraphics[bb=120 505 372 788, width=\textwidth]{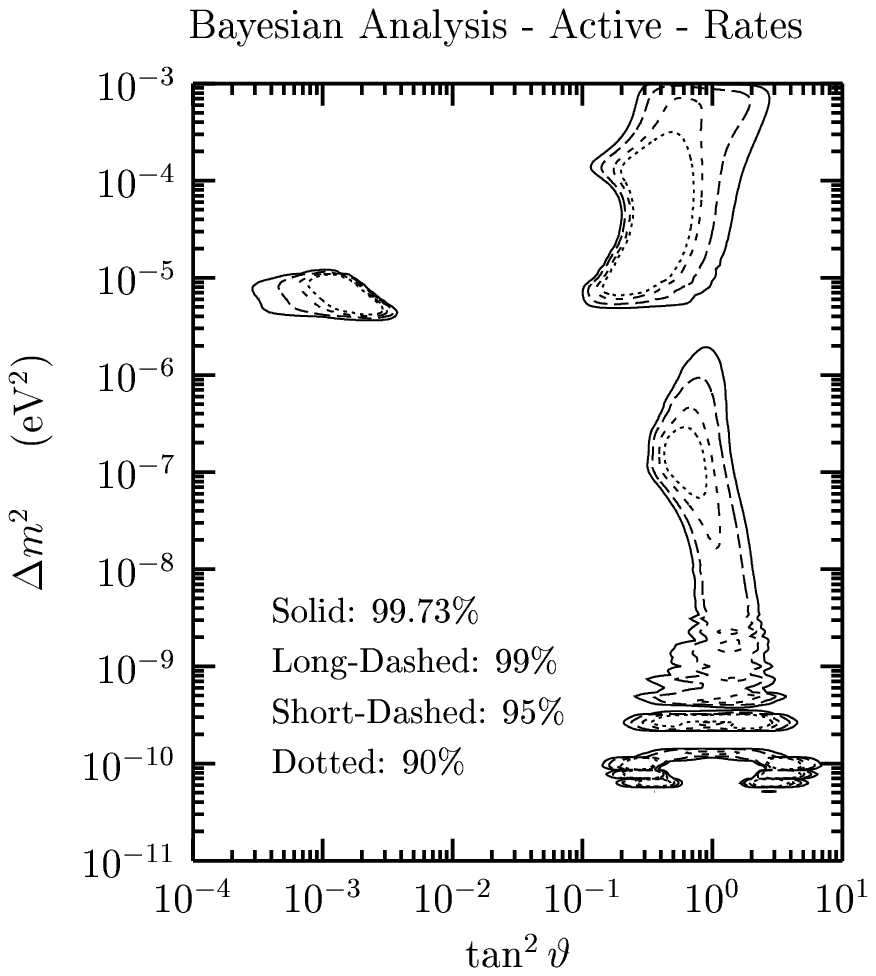}
\caption{
\label{acr-yab}
Credible regions obtained with the Bayesian Rates Analysis
of solar neutrino rates and CHOOZ data
in terms of
$\nu_e\to\nu_a$
transitions.
The dotted, short-dashed, long-dashed and solid
curves enclose credible regions
with, respectively,
90\%,
95\%,
99\%
and
99.73\%
posterior probability to contain the true values of
$\tan^2\!\vartheta$ and $\Delta{m}^2$.
}
}

\FIGURE{
\includegraphics[bb=120 505 372 788, width=\textwidth]{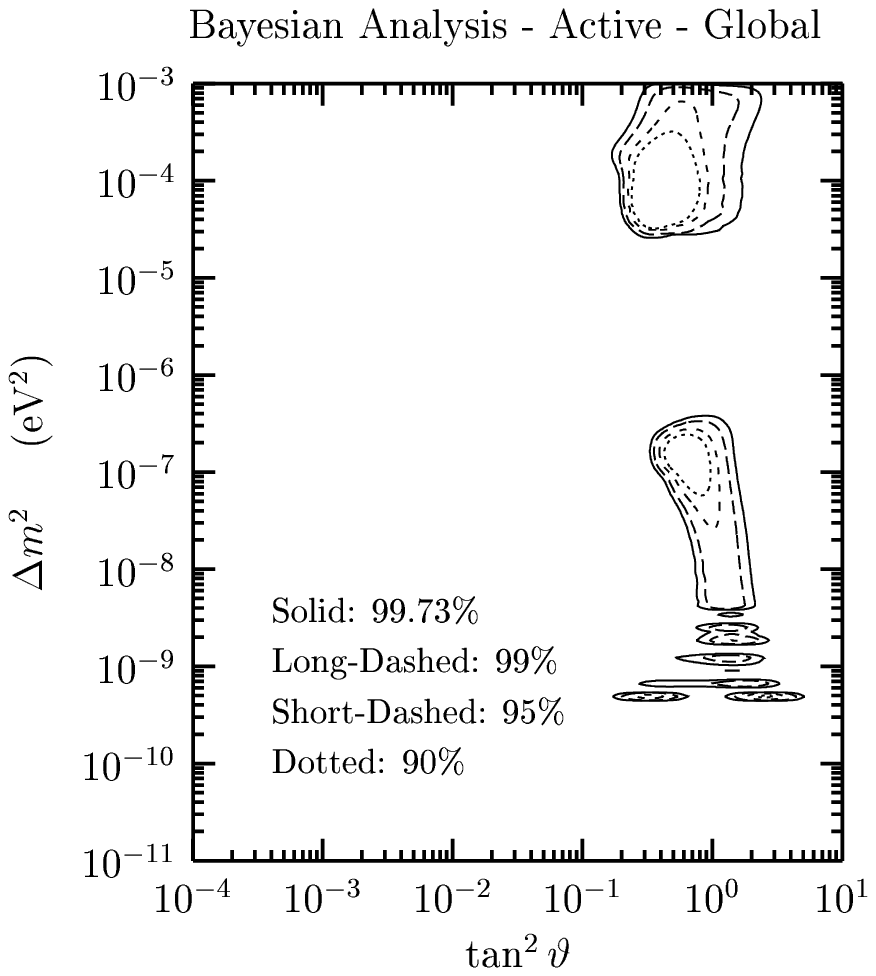}
\caption{
\label{acg-yab}
Credible regions obtained with the Bayesian Global Analysis
of the rates of the Homestake,
GALLEX+GNO+SAGE
and
SNO experiments,
the Super-Ka\-mio\-kan\-de day and night electron energy spectra
and CHOOZ data
in terms of
$\nu_e\to\nu_a$
transitions.
The dotted, short-dashed, long-dashed and solid
curves enclose credible regions
with, respectively,
90\%,
95\%,
99\%
and
99.73\%
posterior probability to contain the true values of
$\tan^2\!\vartheta$ and $\Delta{m}^2$.
}
}

\FIGURE{
\includegraphics[bb=87 424 434 765, width=\textwidth]{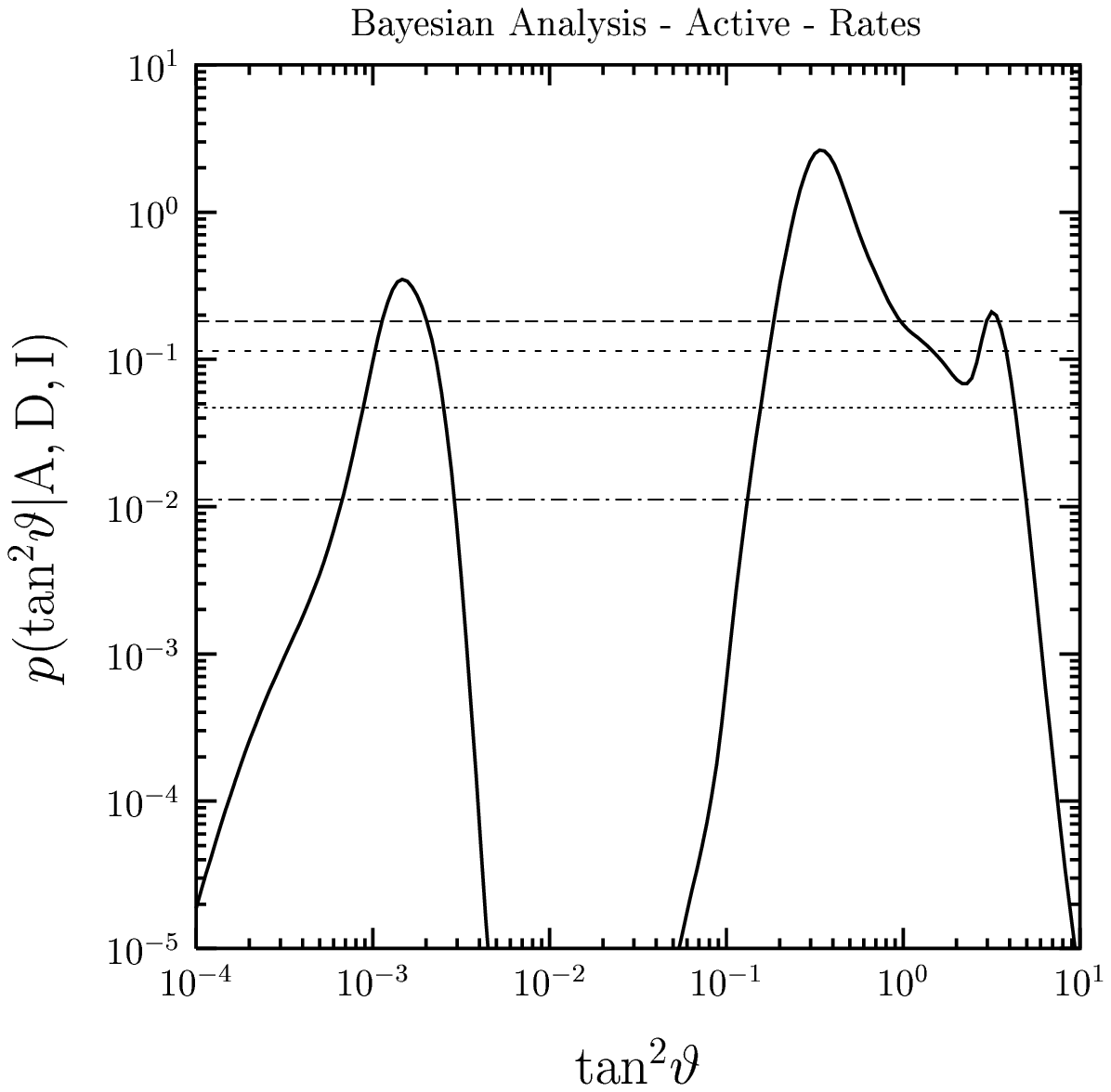}
\caption{
\label{acr-yab-x}
Marginal posterior probability distribution
for $\tan^2\!\vartheta$
obtained with the Bayesian Rates Analysis
of solar neutrino rates and CHOOZ data
in terms of
$\nu_e\to\nu_a$
transitions (solid curve).
The intervals in which
the solid curve lies above the
long-dashed, short-dashed, dotted, and dash-dotted lines
have, respectively,
90\%,
95\%,
99\%
and
99.73\%
posterior probability to contain the true value of
$\tan^2\!\vartheta$.
}
}

\FIGURE{
\includegraphics[bb=87 424 434 765, width=\textwidth]{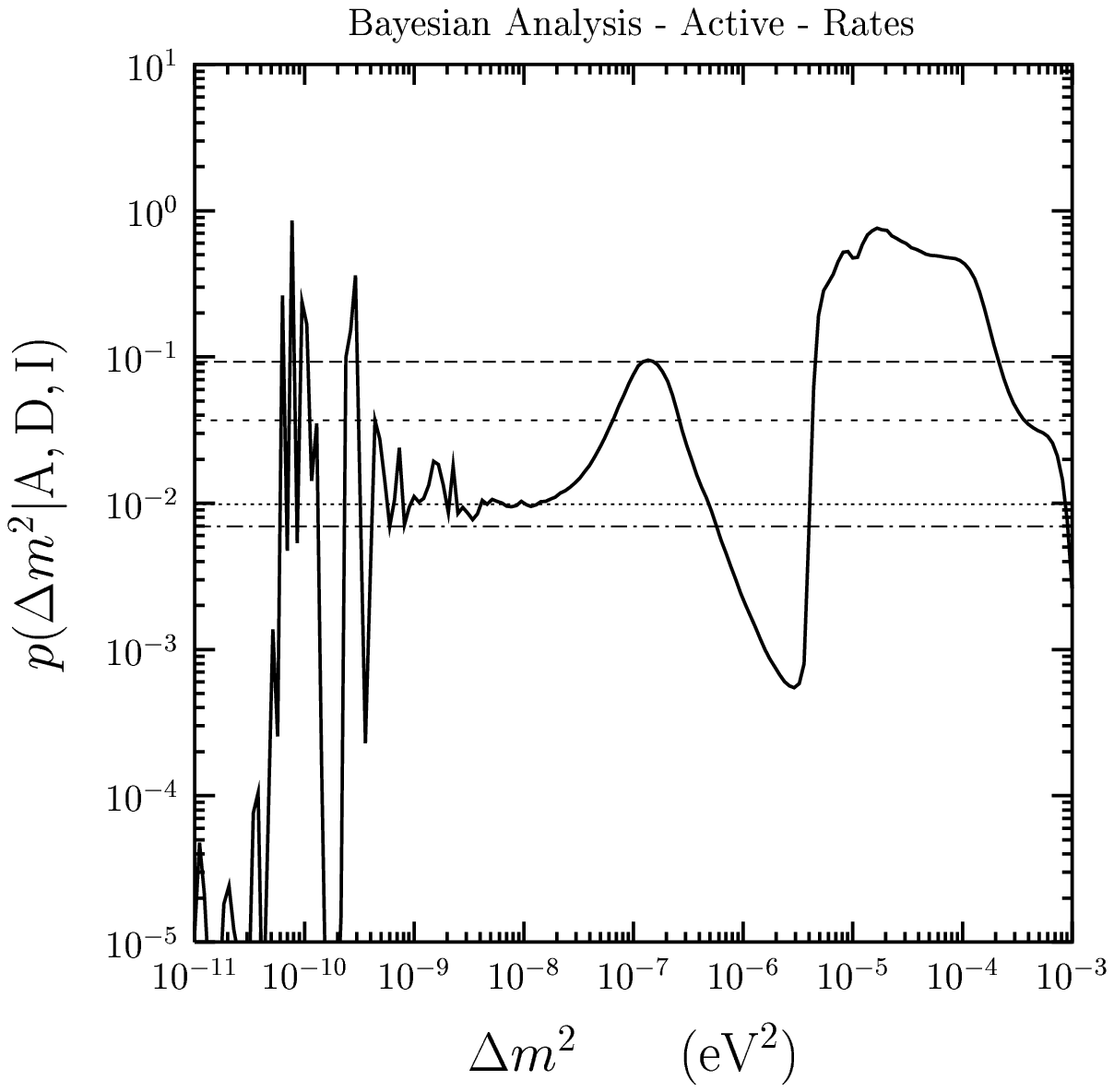}
\caption{ \label{acr-yab-y}
Marginal posterior probability distribution
for $\Delta{m}^2$
obtained with the Bayesian Rates Analysis
of solar neutrino rates and CHOOZ data
in terms of
$\nu_e\to\nu_a$
transitions (solid curve).
The intervals in which
the solid curve lies above the
long-dashed, short-dashed, dotted, and dash-dotted lines
have, respectively,
90\%,
95\%,
99\%
and
99.73\%
posterior probability to contain the true value of
$\Delta{m}^2$.
}
}

\FIGURE{
\includegraphics[bb=87 424 434 765, width=\textwidth]{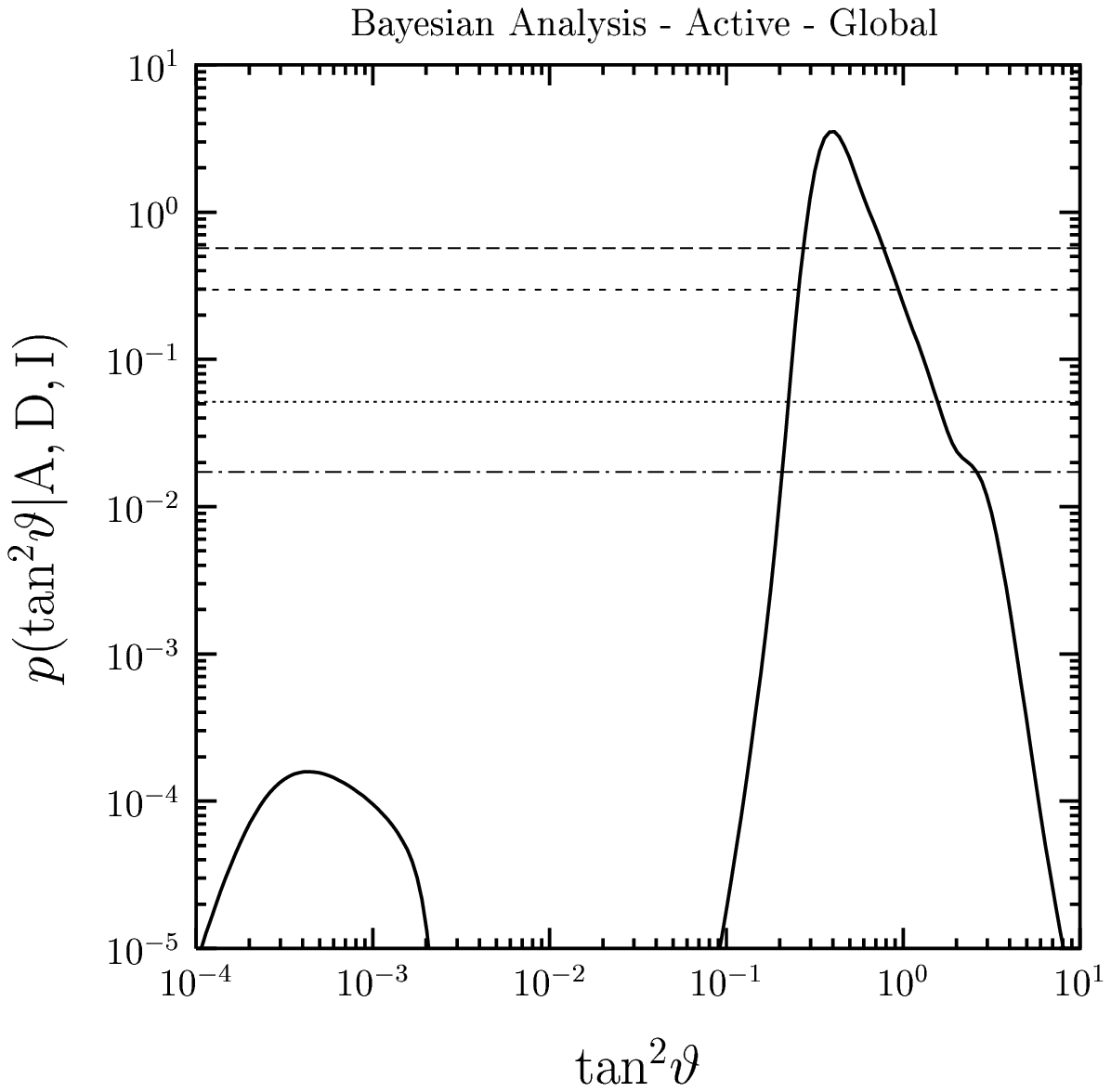}
\caption{
\label{acg-yab-x}
Marginal posterior probability distribution
for $\tan^2\!\vartheta$
obtained with the Bayesian Global Analysis
of the rates of the Homestake,
GALLEX+GNO+SAGE
and
SNO experiments,
the Super-Ka\-mio\-kan\-de day and night electron energy spectra
and CHOOZ data
in terms of
$\nu_e\to\nu_a$
transitions (solid curve).
The intervals in which
the solid curve lies above the
long-dashed, short-dashed, dotted, and dash-dotted lines
have, respectively,
90\%,
95\%,
99\%
and
99.73\%
posterior probability to contain the true value of
$\tan^2\!\vartheta$.
}
}

\FIGURE{
\includegraphics[bb=87 424 434 765, width=\textwidth]{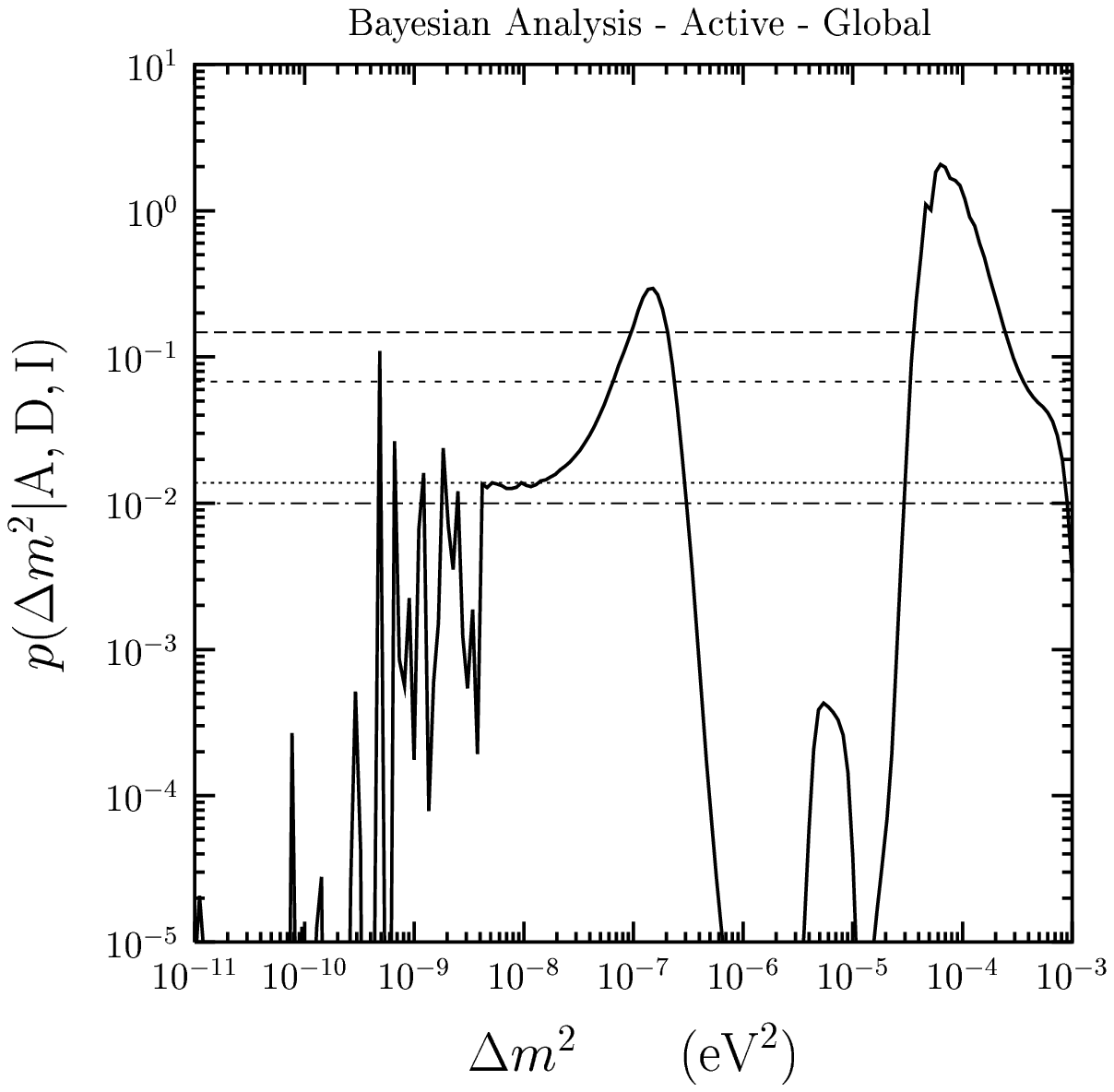}
\caption{
\label{acg-yab-y}
Marginal posterior probability distribution
for $\Delta{m}^2$
obtained with the Bayesian Global Analysis
of the rates rates of the Homestake,
GALLEX+GNO+SAGE
and
SNO experiments,
the Super-Ka\-mio\-kan\-de day and night electron energy spectra
and CHOOZ data
in terms of
$\nu_e\to\nu_a$
transitions (solid curve).
The intervals in which
the solid curve lies above the
long-dashed, short-dashed, dotted, and dash-dotted lines
have, respectively,
90\%,
95\%,
99\%
and
99.73\%
posterior probability to contain the true value of
$\Delta{m}^2$.
}
}

\end{document}